\documentclass[longauth]{aa}  

\usepackage{txfonts}
\usepackage{graphicx}
\usepackage[colorlinks=true,citecolor=blue]{hyperref}
\bibpunct{(}{)}{;}{a}{}{,}

\begin{document} 

   \title{Investigating three Sirius-like systems with SPHERE\thanks{Based on data collected at the European Southern Observatory, Chile (ESO Programs 095.C-0298, 096.C-0241, 097.C-0865, 198.C-0209, 1100.C-0481}}

   \author{
    R. Gratton\inst{1}
       \and
   V. D'Orazi\inst{1}
    \and
    T. A. Pacheco\inst{2}
    \and
    A. Zurlo\inst{3,4,5}
    \and
    S. Desidera\inst{1}
    \and
    J. Mel\'endez\inst{2}
    \and
    D. Mesa\inst{1}
    \and
    R. Claudi\inst{1}
    \and
    M. Janson\inst{6,7}
    \and
    M. Langlois\inst{8,5}
    \and
    E. Rickman\inst{9}
    \and
    M. Samland\inst{6,7}
    \and
    T. Moulin\inst{10}
    \and
    C. Soenke\inst{11}
    \and
    E. Cascone\inst{12}
    \and
    J. Ramos\inst{7}
    \and
    F. Rigal\inst{13}
    \and
    H. Avenhaus\inst{14}
    \and
J.L. Beuzit\inst{5,10}
    \and
B. Biller\inst{7,15}
    \and
A. Boccaletti\inst{16}
    \and
M. Bonavita\inst{1,15}
    \and
M. Bonnefoy\inst{10}
    \and
W. Brandner\inst{7}
    \and
G. Chauvin\inst{10,17}
    \and
M. Cudel\inst{10}
    \and
S. Daemgen\inst{14}
    \and
P. Delorme\inst{10}
    \and
C. Desgrange\inst{8,3}
    \and
N. Engler\inst{14}
    \and
M. Feldt\inst{7}
    \and
C. Fontanive\inst{1,18}
    \and
R. Galicher\inst{16}
    \and
A. Garufi\inst{14,19}
    \and
D. Gasparri\inst{27}
    \and
C. Ginski\inst{20}
    \and
J. Girard\inst{21}
    \and
J. Hagelberg\inst{22}
    \and
S. Hunziker\inst{14}
    \and
M. Kasper\inst{11}
    \and
M. Keppler\inst{7}
    \and
A.-M. Lagrange\inst{10,16}
    \and
J. Lannier\inst{10}
    \and
C. Lazzoni\inst{1}
    \and
H. Le Coroller\inst{5}
    \and
R. Ligi\inst{23,29}
    \and
M. Lombart\inst{17,8} 
    \and
A.-L. Maire\inst{7,24}
    \and
M.R. Mayer\inst{14,25}
    \and
S. Mazevet\inst{28}
    \and
F. M\'enard\inst{10}
    \and
D. Mouillet\inst{10}
    \and
C. Perrot\inst{16,30,31}
    \and
S. Peretti\inst{22}
    \and
S. Petrus\inst{10}
    \and
A. Potier\inst{16}
    \and
D. Rouan\inst{10}
    \and
H.M. Schmid\inst{14}
    \and
T.O.B. Schmidt\inst{16}
    \and
E. Sissa\inst{1}
    \and
T. Stolker\inst{26}
    \and
G. Salter\inst{5}
    \and
A. Vigan\inst{5}
    \and
F. Wildi\inst{22}
}

   \institute{\inst{1} INAF Osservatorio Astronomico di Padova, vicolo dell'Osservatorio 5, 35122, Padova, Italy 
   \email{raffaele.gratton@inaf.it} \and
    \inst{2}Departamento de Astronomia do IAG/USP, Universidade de S\~{a}o Paulo, Rua do M\~{a}tao, 1226, 05508-900, S\~{a}o Paulo, SP, Brazil \and
     \inst{3} N\'ucleo de Astronom\'ia, Facultad de Ingenier\'ia, Universidad Diego Portales, Av. Ejercito 441, Santiago, Chile \and
     \inst{4} Escuela de Ingenier\'ia Industrial, Facultad de Ingenier\'ia y Ciencias, Universidad Diego Portales, Av. Ejercito 441, Santiago, Chile \and
     \inst{5} Aix-Marseille Universit\'e, CNRS, LAM (Laboratoire d'Astrophysique de Marseille) UMR 7326, 13388, Marseille, France\and
     \inst{6} Department of Astronomy, Stockholm University, AlbaNova University Center, 10691 Stockholm, Sweden\and
     \inst{7}Max Planck Institute for Astronomy, K\"onigstuhl 17, D-69117 Heidelberg, Germany\and
     \inst{8} CRAL, UMR 5574, CNRS, Universit\'e Lyon 1, 9 avenue Charles Andr\'e, 69561 Saint Genis Laval Cedex, France\and
     \inst{9} European Space Agency (ESA), ESA Office, Space Telescope Science Institute, 3700 San Martin Drive, Baltimore, MD21218, USA\and
\inst{10} Univ. Grenoble Alpes, CNRS, IPAG, F-38000 Grenoble, France. \and
\inst{11} European Southern Observatory (ESO), Karl-Schwarzschild-Str. 2, 85748 Garching, Germany\and
\inst{12} INAF - Osservatorio Astronomico di Capodimonte, Salita Moiariello 16, 80131 Napoli, Italy\and
\inst{13} NOVA Optical Infrared Instrumentation Group, Oude Hoogeveensedijk 4, 7991 PD Dwingeloo, The Netherlands\and
\inst{14} ETH Zurich, Institute for Particle Physics and Astrophysics, Wolfgang-Pauli-Str. 27, CH-8093, Zurich, Switzerland\and
\inst{15} Institute for Astronomy, University of Edinburgh, EH9 3HJ, Edinburgh, UK\and
\inst{16} LESIA, Observatoire de Paris, Universit\'e PSL, CNRS, Sorbonne Universit\'e, Univ. Paris Diderot, Sorbonne Paris Cit\'e, 5 place Jules Janssen, 92195 Meudon, France\and
\inst{17} Unidad Mixta Internacional Franco-Chilena de Astronom\'{i}a, CNRS/INSU UMI 3386 and Departamento de Astronom\'{i}a, Universidad de Chile, Casilla 36-D, Santiago, Chile\and
\inst{18} Center for Space and Habitability, University of Bern, 3012 Bern, Switzerland\and
\inst{19} INAF - Osservatorio Astrofisico di Arcetri\and
\inst{20} Leiden Observatory, Leiden University, P.O. Box 9513, 2300 RA Leiden, The Netherlands\and
\inst{21} European Southern Observatory, Alonso de Cordova 3107, Casilla 19001 Vitacura, Santiago 19, Chile\and
\inst{22} Geneva Observatory, University of Geneva, Chemin des Mailettes 51, 1290 Versoix, Switzerland\and
\inst{23} INAF - Osservatorio Astronomico di Brera\and
\inst{24} STAR Institute, Universit\'e de Li\'ege, All\'ee du Six Ao\^ut 19c, 4000
Li\'ege, Belgium\and
\inst{25}Department of Astronomy, University of Michigan, 1085 S. University, Ann Arbor, MI 48109, USA\and
\inst{26}Anton Pannekoek Institute for Astronomy, University of Amsterdam, Science Park 904, 1098 XH Amsterdam, The Netherlands\and
\inst{27}INCT, Universidad De Atacama, Calle Copayapu 485, Copiap\'o, Atacama, Chile\and
\inst{28}Laboratoire Univers et Th\'eories, Universit\'e Paris Diderot, Observatoire de Paris, PSL University, 5 Place Jules Janssen, 92195
Meudon, France\and
\inst{29}Universit\'e C\^ote d'Azur, Observatoire de la C\^ote d'Azur, CNRS, Laboratoire Lagrange, Boulevard de l'Observatoire, CS 34229, 06304 Nice Cedex 4, France
\inst{30}Instituto de F\'isica y Astronom\'ia, Facultad de Ciencias, Universidad de Valpara\'iso, Av. Gran Breta\~na 1111, Valpara\'iso, Chile\and
\inst{31}N\'ucleo Milenio Formaci\'on Planetaria - NPF, Universidad de Valpara\'iso, Av. Gran Breta\~na 1111, Valpara\'iso, Chile
}

   \date{Received ; accepted }

 
  \abstract
   {Sirius-like systems are relatively wide binaries  with a separation from a few to hundreds of au; they are composed of a white dwarf (WD) and a companion of a spectral type earlier than M0. Here we consider main sequence (MS) companions, where the WD progenitor evolves in isolation, but its wind during the former asymptotic giant branch (AGB) phase pollutes the companion surface and transfers some angular momentum. They are rich laboratories to constrain stellar models and binary evolution.}
   {Within the SpHere INfrared survey for Exoplanet (SHINE) survey that uses the Spectro-Polarimetric High-contrast Exoplanet REsearch (SPHERE) instrument at the Very Large Telescope (VLT), our goal is to acquire high contrast multi-epoch observations of three Sirius-like systems, HD~2133, HD~114174, and CD-56~7708 and to combine this data with archive high resolution spectra of the primaries, TESS archive, and literature data.}
   {These WDs are easy targets for SPHERE and were used as spectrophotometric standards. We performed very accurate  abundance analyses for the MS stars using methods considered for solar analogs. Whenever possible, WD parameters and orbits were obtained using Monte Carlo Markov Chain (MCMC) methods. }
   {We found brighter $J$\ and $K$\ magnitudes for HD~114174B than obtained previously and extended the photometry down to 0.95~$\mu$m. Our new data indicate a higher temperature and then shorter cooling age ($5.57\pm 0.02$~Gyr) and larger mass ($0.75\pm 0.03$~$M_\odot$) for this WD than previously assumed. Together with the oldest age for the MS star connected to the use of the Gaia DR2 distance, this solved the discrepancy previously found with the age of the MS star. The two other WDs are less massive, indicating progenitors of $\sim 1.3$~$M_\odot$ and $1.5-1.8$~$M_\odot$ for HD~2133B and CD-56~7708B, respectively. In spite of the rather long periods, we were able to derive useful constraints on the orbit for HD~114174 and CD-56~7708. They are both seen close to edge-on, which is in agreement with the inclination of the MS stars that are obtained coupling the rotational periods, stellar radii, and the projected rotational velocity from spectroscopy. The composition of the MS stars agrees fairly well with expectations from pollution by the AGB progenitors of the WDs: HD~2133A has a small enrichment of n-capture elements, which is as expected for pollution by an AGB star with an initial mass $<1.5$~$M_\odot$; CD-56~7708A is a previously unrecognized mild Ba-star, which is also expected due to pollution by an AGB star with an initial mass in the range of $1.5-3.0$~$M_\odot$; and HD~114174 has a very moderate excess of n-capture elements, which is in agreement with the expectation for a massive AGB star to have an initial mass $>3.0$~$M_\odot$}
   {On the other hand, none of these stars show the excesses of C that are expected to go along with those of n-capture elements. This might be related to the fact that these stars are at the edges of the mass range where we expect nucleosynthesis related to thermal pulses. More work, both theoretical and observational, is required to better understand this issue.}


   \maketitle
%

\section{Introduction} \label{sec:Introduction}

Using Sirius as the prototype, binary systems composed of a star with a spectral type earlier than M0 and of a white dwarf (WD) are referred to as Sirius-like \citep{holberg2013}. Here we consider those systems with main sequence (MS) companions. The occurrence of these systems is roughly $\approx$ 8 \% of all known WDs populating the solar surroundings, but this value dramatically drops for distances larger than 25 pc due to observational bias (\citealt{crepp2018} and references therein). In the near-IR ($J$, $H$, and $K-$bands), the MS star is much brighter than the WD because of the larger radius; the contrast is then similar to that of brown dwarfs and/or young planets. For this reason, these targets are preferable science objectives for high-contrast imaging facilities. Interesting enough, in several cases the WD nature of the companion was found as a serendipitous discovery (e.g., HD~27442, \citealt{chauvin2007}; HD~8049, \citealt{Zurlo2013}): These objects were originally proposed to be substellar companions of their host stars, which were included in surveys targeting objects presumed to be young based on activity-related signatures. 

Wide binary systems with a WD companion \citep{Jeffries1996}, which are accessible via direct imaging observations, are great astrophysical laboratories. Fundamental information can be inferred on mass-radius relations, cooling time sequences and the age of the system, nucleosynthesis in asymptotic giant branch (AGB) stars, and more (\citealt{parsons2016}, \citeyear{parsons2017}; \citealt{bacchus2017}; \citealt{crepp2018}). In this respect, Sirius-like systems are simpler than classical Ba-stars (i.e., giants: \citealt{Bidelman1951, McClure1980}) where there may be a further interaction between the two stars when the original secondary climbs up the red giant branch (see e.g., \citealt{Escorza2019}) and where mixing occurs due to first or second dredge-up within this component which altered its surface composition. They are also considered to be a more general case than dwarf Ba- and CH stars \citep{McClure1990, Kong2018} because they also include cases where the modifications in the surface composition of the MS stars may be negligible because of the particular mass of the WD progenitor. In these systems, the two components are separated enough so that they can distinctly evolve, except in the giant phase of the original primary, when mass transfer episodes can occur (see e.g., \citealt{Zurlo2013}). Within Sirius-like systems, particular interest concerns those with a Solar-type companion because the surface abundances of the MS star reflect the composition after the pollution by the secondary and more accurate abundances can be obtained. Previous studies of solar analogs with potential WD companions revealed chemical anomalies in the MS star atmosphere, which are likely as a result of mass transfer from a former AGB companion \citep{Schirbel2015,Desidera2016}. Also, angular momentum transfer from the orbital motion to the MS component occurs in detached systems \citep{Jeffries1996, Matrozis2017}, which shows up as an apparent rejuvenation of the system when using diagnostics related to rotation and activity \citep{Zurlo2013}.

In this work, we consider three Sirius-like systems with a small apparent separation and Solar-type MS stars that were observed as spectro-photometric standards for the SpHere INfrared survey for Exoplanet (SHINE) survey (\citealt{chauvin2017}; \citealt{vigan2020}; Desidera et al., submitted; Langlois et al., submitted). These objects were extracted from the original list of \cite{holberg2013}, and they were selected because it is possible to observe them well with the Spectro-Polarimetric High-contrast Exoplanet REsearch (SPHERE) instrument at VLT: They are accessible from Paranal ($Dec<10$ degree), have a bright MS star ($V<11$), and have a suitable separation (between 0.2 and 0.8 arcsec). Two of them, HD~2133 and CD-56~7708, include a rather hot DA WD that is very bright in the UV \citep{barstow2001}. The MS stars are quite active and appear very young when using activity-related diagnostics. The third system, HD~114174, is a very interesting binary at only $26.38\pm 0.04$~pc \citep{Gaia2018}, containing a cool WD, which is a spectral benchmark for this kind of object and has been discovered through high contrast imaging (\citealt{crepp2013}; \citealt{bacchus2017}). The rather old MS star is considered to be a solar analog and then included in various specific studies providing extremely accurate differential abundances \citep{Chen2003, Ramirez2009, DoNascimento2010, Takeda2010, Ramirez2014}. The binary period is short enough to produce significant radial velocity variations \citep{crepp2013} and astrometric motion so that we may derive constraints on the orbital parameters \citep{bacchus2017}. The cooling age of $7.77\pm 0.24$~Gyr derived by \citet{Matthews2014} (see also \citealt{bacchus2017}) for this WD appears longer than the age of the MS star derived from its position in the color-magnitude diagram ($4.7^{+2.3}_{-2.6}$~Gyr) or from gyrochronology ($4.0^{+0.96}_{-1.09}$~Gyr); if confirmed, this circumstance would require one to substantially rethink the models. We present the spectro-photometric and astrometric data for these three systems gathered in the first four years of the SHINE survey. 
In addition, we present an analysis of high dispersion spectra of the primaries, which are aimed to detect rotation and understand if the composition of the MS star is consistent with the expectations of materials transferred from the WD progenitors.

The paper is organized as follows. In Section~\ref{sec:observations} we present high-contrast imaging observations with SPHERE, as well as complementary high-resolution spectra of the primaries from archives. In Section~\ref{sec:sphere} we describe the reduction and analysis of the SPHERE data. In Section~\ref{sec:ages} we discuss the ages of the systems derived from different methods; these also allow us to infer the initial masses of the WD progenitors. In Section~\ref{sec:primaries} we present the results of the analysis of the high dispersion spectra of the MS stars. The derivation of preliminary orbital parameters for the three systems are given in Section~\ref{sec:orbit}. Finally, the results are discussed and conclusions are drawn in Section~\ref{sec:discussion}.

\section{Observations and data reduction} \label{sec:observations}

\subsection{SPHERE data}

We targeted the three systems as spectro-photometric standards for the SHINE survey (see \citealt{vigan2020} and references therein) with SPHERE located on UT3 at the ESO Very Large Telescope \citep{beuzit2019}. Our observations cover a temporal range between February 2015 $-$ May 2019 for HD~114174, October 2015 $-$ December 2018 for HD~2133, and September 2015 $-$ September 2018 for CD-56~7708 (see the complete observing log reported in Table~\ref{tab:observations}). We acquired data with the typical observing procedure used for the survey (Langlois et al. 2020, submitted), though often the observing sequence was kept rather short and the observations were acquired far from the meridian because of the lower priority given to these targets with respect to the main purposes of the SHINE survey. For this reason and for the occasionally poor weather conditions, some of the observations are of a poor quality and could not be used in this paper. Briefly, the high-contrast imager SPHERE, with the high-order AO system SAXO \citep{fusco2006}, was used with the two infrared channels: the integral field spectrograph IFS \citep{claudi2008} and the dual band imager IRDIS \citep{dohlen2008, vigan2010}. IFS and IRDIS were used in parallel mode; a large part of the observations were performed with SPHERE using the IRDIFS observing mode, that is with IFS operating in the $Y$ and $J$ spectral bands between 0.95 and 1.35 micron and IRDIS exploiting $H2-H3$ narrow band filters (wavelength $H2$=1.59 micron; wavelength $H3$=1.66 micron). However, a fraction of the observations were done in IRDIFS-EXT mode, that is using IFS in the $YH$ mode (wavelength range 0.95-1.65 micron, resolving power R$\approx$30) and IRDIS in $K1-K2$ mode (i.e., 2.09 and 2.25 micron). We acquired the observations in pupil-stabilized mode with an Apodized Lyot Coronagraph with a focal mask having a diameter of 185 mas \citep{boccaletti2008}. We also obtained several on-sky calibrations for each scientific observation: a PSF flux calibration, with the star offset with respect to the coronagraphic mask; centering calibrations, where we obtained satellite images symmetric with respect to the central star by imparting a bi-dimensional sinusoidal pattern to the deformable mirror; and sky calibrations that are important for background subtraction on IRDIS data at long wavelengths. We also acquired parallel data sets. In particular, they include sequences of $H-$band images acquired by the sensor used by the active loop on the tip-tilt mirror that centered the star image on the coronagraphic mask (DTTS). While these are only snapshots, which were taken every 30 sec, they provide a useful monitoring of the fluctuations of transmission and image quality.

Data were reduced using the standard SPHERE pipeline (v. 15.0; \citealt{pavlov2008}), and then by a suite of routines available in the consortium Data Center in Grenoble \citep{delorme2017}. The final output of the data reduction procedure consists of four dimensional datacubes that include spatial (two dimensional), temporal, and wavelength information.

\begin{center}
\begin{table*}[htb]
\caption{Observation log for HD 114174, HD 2133, and CD-567708.}
\label{tab:observations}
\begin{tabular}{lccccccccccr}
\hline
\hline
JD      & Mode  & N$_{\rm DIT}$xDIT & Rotation & Seeing  &      Contrast  &      Sep     & PA & Quality &        d$Y$ &  d$J$ &  d$H_{\rm IFS}$ \\
&  & & &  FWHM & at {\tiny 0.5 arcsec} &        & & & & &  \\
        &      & sec & deg      & arcsec        & mag   & mas   & deg &  & mag   & mag & mag \\
        \hline
& & & & & & & & &  & \\
    &  & & & &  \textbf{HD 114174} &  & & & & & \\ 
& & & & & & & & &  & \\
2457058.37&     IRDIFS& 34x32&  15.52&  1.67&   13.13&  640.5&  172.03& Good&   10.11&  10.04 &       10.26 \\
2457112.20&     IRDIFS& 12x64&  5.97&   0.67&   13.65&  637.2&  171.71& Good&   9.87 &       10.01&  10.26\\
2457146.09&     IRDIFS& 48x64&  23.47&  0.62&   14.82&  635.6&  172.02& Good&   10.07&  10.09&  10.21\\
2457174.05&     IRDIFS& 15x64&  7.96&   1.21&   12.98&  634.1&  172.24& Good&   10.19&  9.92 &       10.24\\
2457180.12&     IRDIFS& 15x32&  0.00&   1.60&   8.49&   635.6&  171.60& Poor&   ----     &         ----   &   ----   \\
2457207.95&     IRDIFS& 68x64&  59.20&  1.78&   15.39&  634.8&  172.00& Poor&   ----     &         ----   &   ----   \\
2457208.04&     IRDIFS& 16x64&  11.87&  2.10&   13.28&  635.5&  172.15& Poor&   ----     &         ----   &   ----   \\
2457791.21&     IRDIFS& 60x32&  3.90&   0.64&   12.64&  601.9&  173.47& Good&   9.96 &       10.23&  10.06\\
2457792.22&     IRDIFS-EXT&     60x32&  4.63&   0.55&   13.09&  600.1&  172.27& Good&   9.84 &       9.97 &  10.17\\
2457831.83&     IRDIFS& 60x64&  6.62&   1.63&   11.74&  604.4&  172.30& Good&   10.29&  10.35&  10.31\\
2457882.15&     IRDIFS-EXT&     32x32&  6.16&   1.01&   14.15&  591.1&  171.88& Good&   9.93 &       10.26&  10.28\\
2458175.16&     IRDIFS& 64x32&  4.10&   0.80&   12.24&  583.2&  173.07& Good&   10.25&  10.29&   ----   \\
2458220.08&     IRDIFS& 64x32&  7.51&   0.67&   12.27&  577.3&  172.42& Good&   10.06&  9.71 &        ----   \\
2458244.01&     IRDIFS& 64x32&  6.99&   0.66&   11.51&  573.9&  172.49& Good&   9.77 &       9.92 &   ----   \\
2458255.96&     IRDIFS& 64x32&  5.53&   0.84&   11.85&  574.4&  172.53& Good&   10.03&  9.72 &        ----   \\
2458549.15&     IRDIFS& 4x32&   0.19&   0.54&   9.45&   559.4&  172.34& Poor&   ----     &       ----     &  ----    \\
2458553.13&     IRDIFS-EXT&     64x32&  4.30&   0.43&   13.19&  559.2&  172.64& Good&   ----     &  10.17&       10.32\\
2458587.08&     IRDIFS& 64x32&  7.38&   0.36&   12.79&  553.4&  172.61& Good&   ----     &  10.03&  ----     \\
2458621.99&     IRDIFS& 64x32&  8.67&   0.74&   12.15&  553.1&  172.58& Good&   ----     &  10.08&  ----     \\
\hline
& & & & & & & & &  & \\
    &  & & & &  \textbf{HD 2133} &  & & & & & \\ 
& & & & & & & & &  & \\
2457320.15 &    IRDIFS  &63x64 & 20.56& 2.03&   11.85&  675.0&  49.31 & Good &       7.50 &  7.74 &  8.10\\
2457322.15 &    IRDIFS-EXT      &15x64 & 5.16&  1.85&   11.07&  673.7&  49.28 &       Good &  7.57 &  7.72 &  8.04\\
2457354.13 &    IRDIFS  &14x32 & 2.42&  1.59&   10.03&  676.1&  49.59 & Good &       7.32 &  7.58 &  7.95\\
2457552.33 &    IRDIFS-EXT      &13x64 & 4.59&  0.85&   11.31&  676.5&  49.29 &       Good &  7.35 &  7.54 &  7.97\\
2457647.05 &    IRDIFS  &20x64 & 11.74& 1.03&   11.10&  677.5&  49.03 & Good &       7.67 &  7.88 &  8.04\\
2457710.10 &    IRDIFS  &16x64 & 5.13&  0.78&   11.61&  678.6&  49.26 & Poor &       ----     &  ----     &  ----    \\              
2457734.10 &    IRDIFS  &5x64  & 11.48& 2.55&    9.02&  689.0&  49.60 & Poor &       ----     &  ----     &  ----    \\      
2457735.03 &    IRDIFS  &4x64  & 1.03&  2.08&   10.47&  679.8&  49.18 & Poor &       ----     &  ----     &  ----   \\       
2458061.01 &    IRDIFS-EXT      &11x64 & 3.37&  1.01&   11.30&  687.5&  49.50 &       Good &  7.37 &  7.64 &  8.02\\
2458087.03 &    IRDIFS-EXT      &16x64 & 5.30&  0.69&   12.19&  683.6&  49.48 &       Good &  7.47 &  7.74 &  8.04\\
2458123.02 &    IRDIFS-EXT      &16x64 & 4.87&  1.21&   11.47&  684.6&  49.36 &       Good &  7.42 &  7.64 &  8.01\\
2458468.02 &    IRDIFS-EXT      &16x64 & 5.18&  0.83&   11.53&  688.7&  49.45 &       Good &  7.85 &  8.07 &  8.35\\
\hline
& & & & & & & & &  & \\
    &  & & & &  \textbf{CD-56~7708} &  & & & & & \\ 
& & & & & & & & &  & \\
2457289.05&     IRDIFS  & 5x64  & 1.72& 1.35&   8.46 &  ----     &       ----     & Poor&        ----    &----         &   ---- \\
2457562.31&     IRDIFS-EXT      & 2x32  & 0.00& 0.56&   11.56&  283.4&  123.64& Poor&   ----    &----         &  ----       \\
2457675.98&     IRDIFS-EXT      & 25x32 & 5.75& 0.65&   11.04&  280.5&  124.72& Good&   5.74&   6.14  &      6.58    \\
2457872.22&     IRDIFS  & 34x32 & 3.90& 0.52&   10.62&  284.0&  123.87& Good&   5.79&   6.11  &      6.42    \\
2457872.42&     IRDIFS-EXT      & 32x32 & 6.96& 0.62&   11.87&  291.0&  123.55& Good&   5.78&   6.15  &      6.51    \\
2457906.28&     IRDIFS-EXT      & 32x32 & 7.65& 0.50&   12.00&  287.7&  124.30& Good&   5.74&   6.11  &      6.48    \\
2458024.10&     IRDIFS  & 47x32 & 7.23& 0.93&   10.77&  293.3&  124.23& Good&   5.72&   6.05  &----          \\
2458379.03&     IRDIFS-EXT      & 32x32 & 7.43& 0.51&   11.00&  291.0&  123.28& Good&   5.62&   6.01  &      6.40    \\
\hline
\end{tabular}
\end{table*}
\end{center}

\subsection{ High resolution spectroscopic data }

High-resolution, high-quality spectra were exploited in order to derive projected rotational velocity, metallicity and carbon, and heavy-element abundances ($s$-process tracers) for the primary stars (see details in Section~\ref{sec:primaries}). Spectroscopic properties of HD~114174A were investigated through HARPS ($\lambda$ 378 - 691 nm, R = 115,000; \citealt{mayor2003}) spectra, which were gathered under program 188.C-0265 (PI: Mel\'endez). The total sample includes 76 spectra, which were observed between 2012 and 2017, with a signal-to-noise ratio (S/N) per spectra ranging from 63 to 163 per pixel at 550 nm. From the ESO archive, we retrieved the FEROS \citep{kaufer99} spectra for HD 2133, which grants a spectral coverage from $\sim$ 350 nm to $\sim$ 950 nm with a nominal resolution of R=48,000. Observations were acquired on June 23, 2012 under program 089.D-0097 (PI: Helminiak); the median S/N per pixel across the orders is 96 per individual spectra, and we combined eight of the best spectra.
Finally, as for CD-56 7708, we analyzed four HARPS spectra available through the ESO archive. Observations were carried out on June 10, 2011 (program: 087.D-0012, PI: Helminiak); the S/N per pixel at 550 nm of the individual spectra is 53. For consistency, the HARPS spectra of HD 114174A and CD-56 7708 were analyzed relative to a solar spectrum also observed with HARPS, and the FEROS spectrum of HD 2133 was analyzed relative to a solar spectrum acquired with FEROS.

The normalization of the spectra of the three stars was done in the same way as with the respective HARPS/FEROS reference solar spectra, using an interval of 10 \AA\  around each line of interest. HD~114174 has a small $v~\sin{i}$, hence the solar spectrum could be directly compared for a differential analysis. However, the other two stars, HD 2133 and CD-56 7708, show broadened line profiles, and therefore before the normalization we broadened the solar spectrum to 25 km s$^{\rm -1}$ (which is close to the $v~\sin{i}$ of those two sample stars), thus providing an appropriate reference for a differential analysis.

\section{Analysis of SPHERE data and results} \label{sec:sphere}

\begin{figure*}[hbt!]
  \centering
  \includegraphics[width=\textwidth]{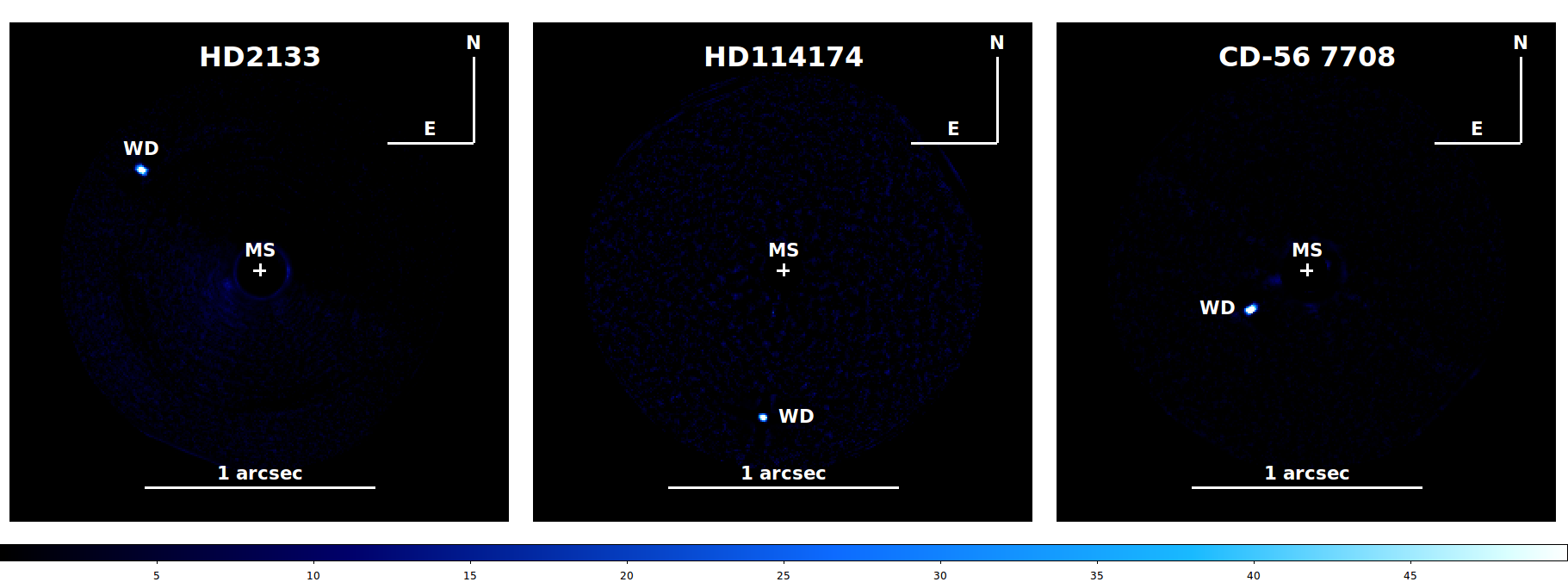}
  \caption{S/N maps obtained from SPHERE IFS data for the three systems. The position of the MS and the WD stars are labeled; it is important to note that the MS star is behind the coronagraphic mask and is not visible in these images, even though it is much brighter than the WD in the near infrared.}
  \label{fig:image}
\end{figure*}

The WDs are clearly seen in the SPHERE images (see Figure~\ref{fig:image}). In the following we report on the results obtained from the analysis of these data.

\subsection{WD photometry}

Photometry of faint companions in the SHINE survey is described in Langlois et al. (2020), where errors are also discussed. The methods used there are based on the SpeCal routines \citep{Galicher2018}. Here we will discuss those features that are relevant for the present cases, where companions are brighter than the typical companions considered for the SHINE survey. We only considered good observations for photometry, while even poor images provided useful data for astrometry. According to our definition, good observations are those that satisfy two criteria. First, the flux measured by the DTTS during the observation agrees with expectation for the MS star (within 0.4 mag). The expectation considers a target magnitude
in the $H-$band, airmass, and DTTS exposure time. The zero-point constant was determined as a modal value using a large set of observations. Second, the flux measured on the PSF calibration agrees with the expectation for the target magnitude (within 0.25 mag). In this case the expectation was obtained as above, but for the $J-$band rather than the $H-$band.

We mostly used IFS data in this paper; IRDIS data were only used to provide magnitudes at longer wavelengths. We derived a contrast between the WD and the star at each wavelength using the negative planet method \citep{Bonnefoy2011, Zurlo2014}: We inserted a negative PSF (obtained from the appropriate calibration) close to the position of the WD, and we optimized its position and intensity by minimizing the r.m.s. scatter of the signal in individual pixels in a $5\times 5$ area centered on the approximate position using an amoeba downhill routine.

The contrast in the different bands was obtained using images that are, on average, over the band-pass of the different filters on the individual IFS channels. The $y-$ band considered here is obtained from the IFS data and it is in the Pan-STARRS1 photometric system \citep{Tonry2012} and it is centered at 0.96~$\mu$m. The $Y-$band is from 1.0 to 1.1~$\mu$m and the $J-$band is similar to the 2MASS one. The $H_{\rm IFS}-$band only covers the range of $1.50-1.65~\mu$m because of limitations in the spectral range covered by IFS and it is then at shorter wavelengths than the 2MASS $H-$band.

Table~\ref{tab:photo} summarizes the average contrasts and the r.m.s. scatter around the mean values in the different bands over all good observations obtained in this way, while Table~\ref{tab:photo2} gives the final results for both the MS and the WD in the three systems. We notice that our estimate of d$J=10.05\pm 0.05$ mag for the contrast between the two components of HD~114174 makes the WD brighter at short wavelengths than found by \citet{crepp2013} (d$J=10.48\pm 0.11$~mag) and \citet{bacchus2017} (d$J=10.33\pm 0.24$; in this second case, however, the result is not really discrepant as it is within the error bars). The contrast in the $y-$band of $9.88\pm 0.05$~mag is fully consistent with the value of d$Y>8.2\pm 0.7$~mag obtained by \citet{Matthews2014}. As we see in Section~\ref{sec:ages}, our new photometry does not require the adoption of a very low temperature (from the brightness in the $L-$band: \citealt{Matthews2014}) as previously found. 

\begin{table*}[htb]
\caption{Average and r.m.s. values for contrasts in different bands for the three WDs (in mag).}
\centering
\begin{tabular}{lcccccccccccc}
\hline
\hline
Star &  Nobs &  d$y$  & rms($y$) & d$Y$ & rms($Y$) & d$J$ & rms($J$) & d$H_{\rm IFS}$ & rms($H_{\rm IFS}$) & rms($IRDIFS$)& rms($IRDIFS-EXT$) \\
\hline
HD~2133     &  9 &  7.15 & 0.05 & 7.50 & 0.17 &  7.73 & 0.16 &  8.06 & 0.12 & 0.04 & 0.06 \\
HD~114174       & 16 &  9.88 & 0.09 & 10.03 & 0.16 & 10.05 & 0.19 & 10.23 & 0.08 & 0.22 & 0.16 \\
CD-56~7708  &  6 &  5.44 & 0.05 & 5.73 & 0.06 &  6.10 & 0.05 &  6.48 & 0.07 & 0.03 & 0.05 \\
\hline
\end{tabular}
\label{tab:photo}
\end{table*}

\begin{table*}[htb]
\caption{MS star and WD photometry.}
\centering
\begin{tabular}{lcccccc}
\hline
\hline
Parameter   & HD~2133&Ref&  HD~114174    &Ref&CD-56~7708\\
\hline
\multicolumn{7}{c}{Main sequence star}\\
\hline
$V$ (mag)              & $9.62 \pm 0.02$  &     1 & 6.80             & 4 & 10.6  & 1 \\
$J$ (mag)              & $8.572\pm 0.023$ &     2 & $5.613\pm 0.026$ & 2 & $9.308\pm 0.027$      & 2 \\
$H$ (mag)              & $8.35\pm 0.021$  &     2 & $5.312\pm 0.027$ & 2 & $8.965\pm 0.026$      & 2 \\
$K$ (mag)              & $8.298\pm 0.026$ &     2 & $5.202\pm 0.023$ & 2 & $8.835\pm 0.019$      & 2 \\
Parallax (mas)         & $7.64\pm 0.03$   &     3 & $37.91\pm 0.05$  & 3 & $7.64\pm 0.12$        & 3 \\
PM RA (mas/yr)         & 2.84             &     3 & 85.36            & 3 & 22.52  & 3 \\
PM Dec (mas/yr)        & -13.79           &     3 & -680.26          & 3 & -38.15 & 3 \\
Spectral Type          & F7V              & 1 & G5IV-V           & 4 & G5V    & 1 \\
$T_{\rm eff}$ Gaia (K) &        6059          & 3 & 5726             & 3 & 5510   & 3 \\
\hline
\multicolumn{7}{c}{White dwarf}\\
\hline
$V$ (mag)   & 15.6                & 1 &                 &   &  14.7           & 1 \\
$y$ (mag)   & $16.40\pm 0.05$ & 5 & $16.20\pm 0.05$ & 5 &
$15.46\pm 0.05$ & 5 \\
$J$ (mag)   & $16.30\pm 0.04$ & 5 & $15.66\pm 0.06$     & 5     & $15.41\pm 0.02$ & 5 \\
$H_{\rm IFS}$ (mag)   & $16.41\pm 0.02$ & 5 & $15.54\pm 0.03$   & 5     & $15.44\pm 0.03$ & 5 \\
$K$ (mag)   & 17.18                   & 5 & $15.22\pm 0.02$     & 5     &                     &   \\
$L$ (mag)       &                 &   & $15.30\pm 0.16$ & 4 &                     &   \\
$M_V$ (mag) & 10.02               & 5 &                 &   &  9.12           & 5 \\
$M_y$ (mag) & $10.82\pm 0.05$ & 5 & $14.09\pm 0.05$     & 5     & $9.87\pm 0.05$  &        5 \\
$M_J$ (mag) & $10.72\pm 0.04$ & 5 & $13.55\pm 0.06$     & 5     & $9.82\pm 0.01$  &        5 \\
$M_H$ (mag) & $10.83\pm 0.02$ & 5 & $13.44\pm 0.03$     & 5     & $9.86\pm 0.02$  &        5 \\
$M_K$ (mag) & 11.61           & 5 & $13.11\pm 0.02$     & 5     &                 &   \\ 
$M_L$ (mag) &                 &   & $13.08\pm 0.16$ & 4 &                         &   \\
\hline
\end{tabular}
\label{tab:photo2}

References: 1: \citet{holberg2013}; 2: \citet{Cutri2003}; 3: \citet{Gaia2016, Gaia2018}; 4: \citet{bacchus2017} ; 5 This paper
\end{table*}



\begin{figure*}[hbt!]
  \centering
  \includegraphics[width=\textwidth]{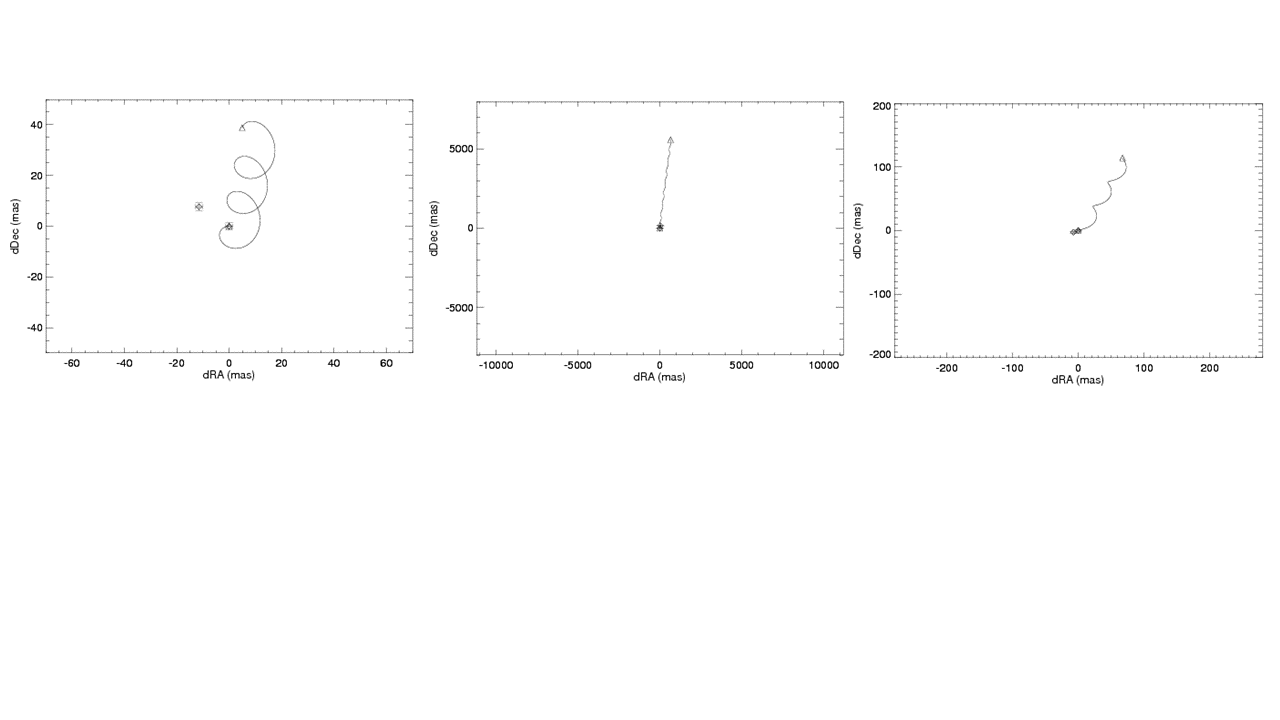}
  \caption{Comparison between the observed relative motion of the two components of HD~2133 (left panel), HD~114174 (central panel), and CD-56~7708 (right panel). Square with error bars are observations at first and last epoch; triangles are predictions for a background object. The lines represent the reflex of the combination of the parallactic and proper motion of the MS star.}
  \label{fig:motion}
\end{figure*}

\subsection{WD astrometry}

Astrometry with SPHERE is presented in \citet{Maire2016} and Langlois et al. (2020, submitted). For our stars, typical astrometric accuracy is about 1.7 mas in both coordinates and is mainly due to uncertainties in star centering. The much shorter separation, and then the base used to define the PA, explains the larger scatter in PA for CD-56~7708. The impact of uncertainties on the scale (typically $<0.05$\%) is negligible, while the definition of true north has a significant impact on the final astrometric errors. Errors involved in centering the companion can be estimated by comparing results obtained by different methods; it turns out to be about 1.0 mas for all three stars that are considered here, irrespective of the very different contrast. We attribute most of the remaining error to the determination of the precise location of the center of the stellar image due to small movements (still of about 1 mas) of the image in between the centering calibration and the observations; uncertainties in these two quantities thoroughly explain the error in PA as well. We note that the specification for centering the star on the coronagraph when designing the tip-tilt system for the SPHERE coronagraph was one mas along both coordinates, so that the error agrees with expectations.

The association of the WD with the MS star could simply be derived by a statistical argument  because it is extremely unlikely to find such a bright MS star simply projected randomly so close to the WD. However, astrometry clearly indicates that the WDs are physically associated with the MS stars. This is illustrated in Figure~\ref{fig:motion}, which compares the observed relative motion of the WDs with respect to the main sequence stars with that expected for a field background star. None of the three companions observed move consistently with the expectation for a background star.

When combined with literature data, we detected orbital motion for all three objects (see Table~\ref{tab:observations}). In the case of HD~114174, we may describe the motion as a linear trend in separation with no clear change in PA since the discovery (about 6 years). This suggests a highly eccentric orbit seen nearly edge on (see \citealt{bacchus2017}). We come back to the constraints on the orbits that can be obtained from our data in Section~\ref{sec:orbit}.

\section{Ages of the stars} \label{sec:ages}

The general evolutionary scenario we consider for these systems is similar to that considered by \citet{Zurlo2013} for the case of HD~8049. Briefly, we start from a relatively wide binary system (e.g., separation between a few to some tens of au) composed of a star in the mass range 1-8~$M_\odot$\ and a smaller mass companion; these systems are very common because a large fraction of the stars in the 1-8~$M_\odot$\ mass range have a similar companion \citep{Raghavan2010, Moe2017} and this range of separations is close to the peak of the binary distribution \citep{Duquennoy1991, Raghavan2010}. Given the large separation, the two components evolve essentially autonomously. However, when the original primary star becomes an AGB star, a small amount of mass and a rather significant amount of angular momentum, from the binary orbit, is transferred on the secondary essentially through the red giant wind. This causes some alteration to the surface chemical composition and a considerable spin up in the secondary, possibly at a velocity comparable to the one the secondary had when at the zero age MS; the reason for this similarity is that in both cases, angular momentum accretion is limited by the stability of the star \citep{Jeffries1996, Matrozis2017}. Then when the original primary becomes a WD, the system becomes wider due to the mass loss by the original primary, and the MS star starts to slow down again due to the angular momentum loss related to a magnetic driven wind, as in single stars. In this scenario, we expect different ages to be obtained depending on the different diagnostics. The comparison of the position in the color magnitude diagram of the MS star with isochrones essentially gives the system age. The position of the WD along the cooling sequence gives the time elapsed since the mass transfer episode. A similar age is provided by gyrochronology applied to the MS star, as well as by other age indicators related to activity (e.g., X-ray luminosity). Lithium is expected to be very low or absent in the atmosphere of the MS star (this agrees with observations for similar systems: see e.g., \citealt{Allen2006a}). The difference between the isochrone as well as the WD and gyrochronology age gives the pre-WD lifetime of the original primary, which can be used to derive its original mass by comparison with evolutionary models. This can also be compared with initial-final mass relations of WDs to determine the current mass of the WD. On the other hand, if dynamical masses are available, they can be used to further constrain the initial-final mass relation for the WD.

Depending on the quality of available data, there may be considerable redundancy, which can be used both to confirm this scenario and to further constrain general data (e.g., the mentioned initial-final mass relation for WD, as done in the case of Sirius - see e.g., \citealt{Cummings2018} - or the calibration of gyrochronological ages). The possible discrepancy found by \citet{Matthews2014} for HD~114174 makes this analysis relevant. In the rest of this subsection, we consider some of these points. The results are shown in Table~\ref{tab:ages}.

\begin{table*}[htb]
\centering
\caption{Ages and relevant parameters.}
\begin{tabular}{lcccccc}
\hline \hline
Parameter                        & HD~2133&Ref&  HD~114174    &Ref&CD-56~7708&Ref\\
\hline
\multicolumn{7}{c}{MS star}\\
\hline
Age isochrones (Gyr)             & $4.8\pm 1.2$ & 5 & $4.7\pm 2.3$ & 4 & $2.6\pm 2.2$ & 5 \\
Age isochrones (Gyr)             &       &   & $6.4\pm 0.7$ & 7 &           &  \\
Age isochrones (Gyr)             &       &   & $7.4\pm 2.0$ & 5 &           &  \\
Mass isochrones ($M_\odot$)      & $1.12\pm 0.03$ & 5 & $0.982\pm 0.014$ & 5 & $1.05\pm 0.03$ & 5 \\
Radius isochrones ($R_\odot$)    & $1.29\pm 0.03$ & 5 & $1.059\pm 0.021$ & 5 & $0.99\pm 0.03$ & 5\\
Rotational period (d)            & 1.316 & 5 &  34.6    & 5 &   2.326   & 5\\
$b_p-r_p$                        & 0.735 & 3 &   0.830      & 3 &   0.897   & 3\\
Age gyrochronology (yr)          & $1.5\times 10^7$ & 5 &   $5.8\times 10^9$      & 5 &   $1.8\times 10^7$ & 5\\
\hline
\multicolumn{7}{c}{White dwarf}\\
\hline
Spectral type                    & DA1.9 & 1 &              &   &   DA1.0   & 1\\
$T_{\rm eff}$ (literature) (K)   & 26800 & 1 &   3810       & 4 &   49460   & 1\\
$T_{\rm eff}$ (literature) (K)   & $29724\pm 158$ & 8 & & & $49037\pm 263$ & 8 \\
$T_{\rm eff}$ (MCMC) (K) & $24569\pm 49$& 6 &$5890\pm 270$& 6 & $42103\pm 4200$ & 6\\
Mass (literature) ($M_\odot$)    &  $0.40\pm 0.14$& 8 & & & $0.524\pm 0.04$& 8 \\
Mass (MCMC) ($M_\odot$) & $0.42\pm 0.00$ &  6 &$0.75\pm 0.03$ & 6 & $0.58\pm 0.04$ & 6\\
Cooling Age (MCMC) (yr) & $(2.03\pm 0.01)\times 10^6$& 6 &$(5.57\pm 0.02)\times 10^9$& 6 & $(4.04 \pm 0.05)\times 10^6$ & 6\\
\hline
\end{tabular}

References: 1: \citet{holberg2013}; 2: \citet{Cutri2003}; 3: \citet{Gaia2016, Gaia2018}; 4: \citet{bacchus2017}; 5: This paper; 6: This paper using \citet{Bergeron1995} pure hydrogen cooling sequences; 7: \citet{Tucci2016}; 8: \citet{Joyce2018}
\label{tab:ages}
\end{table*}

\subsection{Main sequence ages}


The MS star gives information about the age of the HD~114174 system. \citet{crepp2013} obtained an age of $4.7\pm 2.3$~Gyr from isochrones fitting in the color-magnitude diagram. \citet{Tucci2016} derived an older age of $6.4\pm 0.7$~Gyr from a fit with the Yonsey-Yale isochrones in the $T_{\rm eff}-\log{g}$ diagram. We notice that the Gaia parallax of HD~114174 is about 6\% smaller than the Hipparcos value. This makes the MS star intrinsically brighter (by 0.13 mag) and older (by about 2 Gyr) than considered by \citet{crepp2013}. On the other hand, a further constraint on the age of HD~114174 can be obtained by considering that the kinematics (galactic orbit eccentricity of 0.37 and maximum height over the galactic plane of 0.06 kpc: \citealt{Mackereth2018}) and chemical composition ([Fe/H]=$0.016\pm 0.037$, [Si/Fe]=-0.005; [Ca/Fe]=0.023: \citealt{Ramirez2009}) support its membership to the thin disk population. Recent estimates of the age of the thin disk are $8.1\pm 0.6$~Gyr (\citep{Fuhrmann2011}, $<9$~Gyr \citep{DiMatteo2019}, and $7.5\pm 1.2$~Gyr \citep{Kilic2017}. From these determinations, we can set an upper limit of 8.7 Gyr on the age of HD114174. Using this prior, we used the PARSEC isochrones \citep{Bressan2012} models and the bayesian PARAM interface\footnote{http://stev.oapd.inaf.it/cgi-bin/param} to estimate an age of $7.4\pm 2.0$~Gyr, a mass of $1.12\pm 0.03$~$M_\odot$, and a radius of $1.29\pm 0.03$~$R_\odot$\ for HD~114174. The new value for the age is very consistent with that by \citet{Tucci2016}. The quoted uncertainty is the one provided by the PARAM web interface. There is a little inconsistency with the adopted prior due to the fact that the true error bars are asymmetric, that is to say more extended at younger ages because of the slower evolutionary timescales, but they are approximated with a unique value in the output.

We proceeded in the same way, but with no prior, in order to derive the ages, masses, and radii of HD~2133 and CD-56~7708. We obtained ages of $4.8\pm 1.2$ and $2.6\pm 2.2$ Gyr, respectively. Considering the moderately young age of these two systems, there is no impact when using a prior whether or not the thick disk ages are excluded.

\subsection{Gyrochronology ages}

\begin{figure}[hbt!]
  \centering
  \begin{tabular}{cc}
  \includegraphics[width=0.235\textwidth]{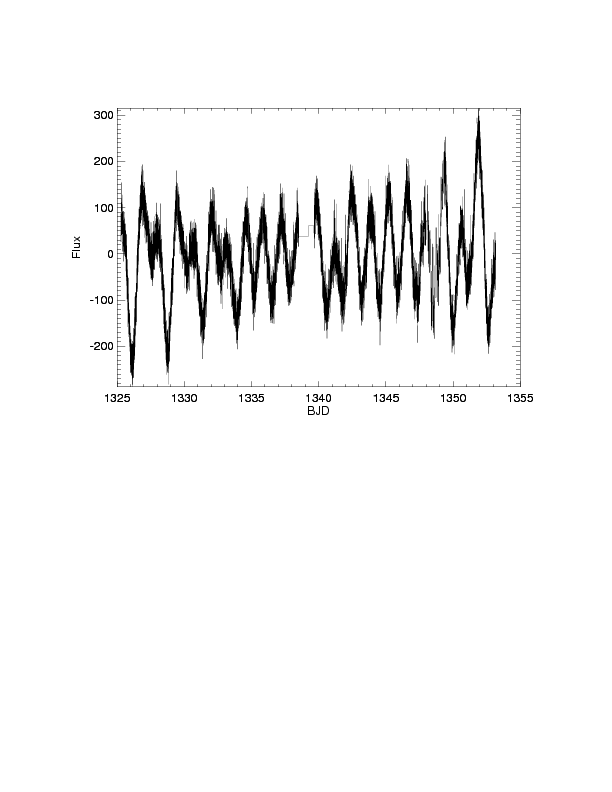}   & 
  \includegraphics[width=0.235\textwidth]{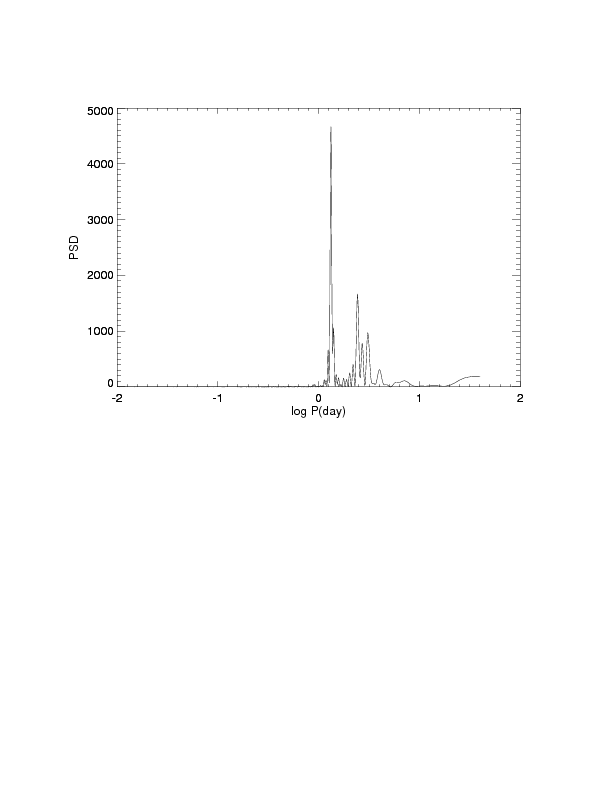}  \\
  \includegraphics[width=0.235\textwidth]{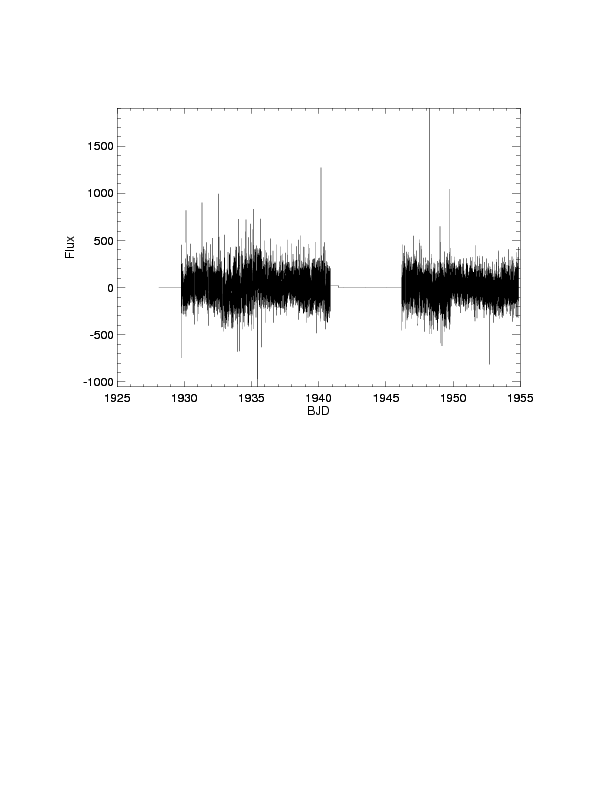}   &
  \includegraphics[width=0.235\textwidth]{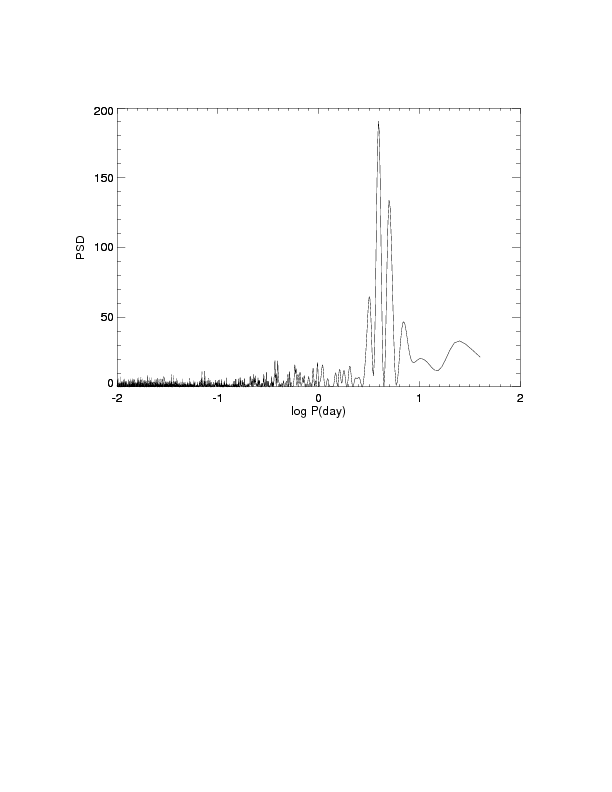} \\
  \includegraphics[width=0.235\textwidth]{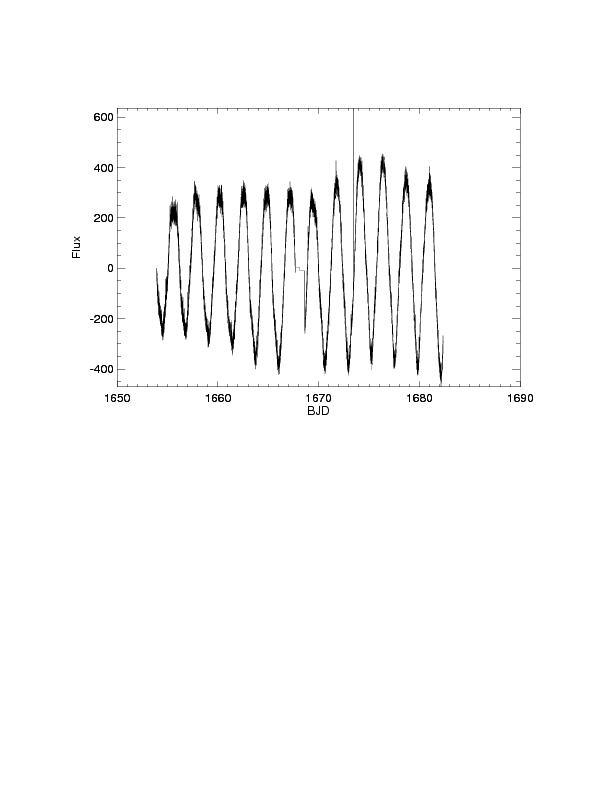}   & 
  \includegraphics[width=0.235\textwidth]{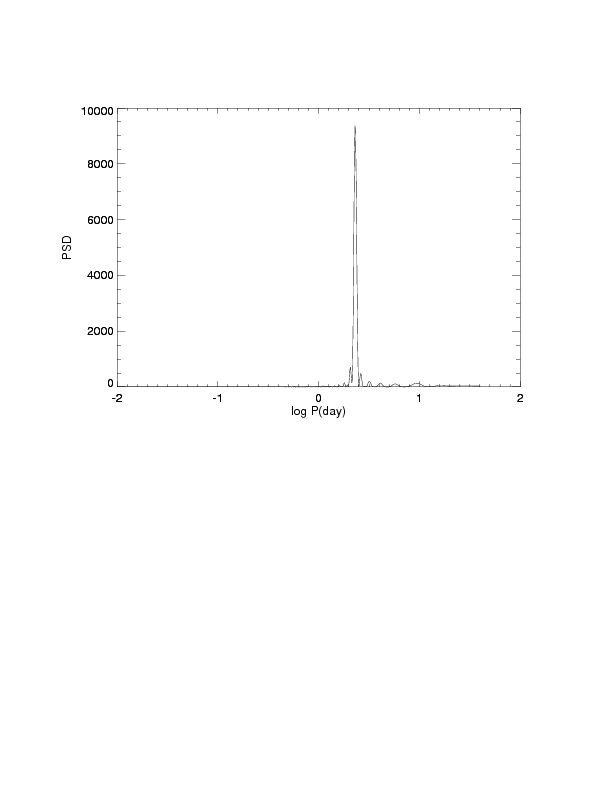}  \\
  \end{tabular}
  \caption{TESS light curves (left) and Scargle periodograms (right) for the three program stars. Upper row: HD2133. Middle row: HD114174. Lower row: CD-56 7708.}
  \label{fig:tess}
\end{figure}

\begin{figure}[hbt!]
  \centering
  \includegraphics[width=0.235\textwidth]{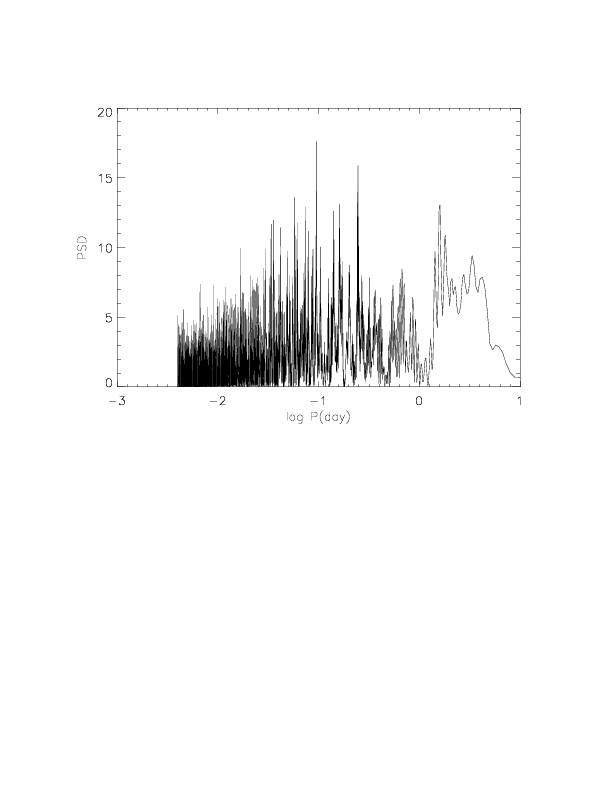}
  \includegraphics[width=0.235\textwidth]{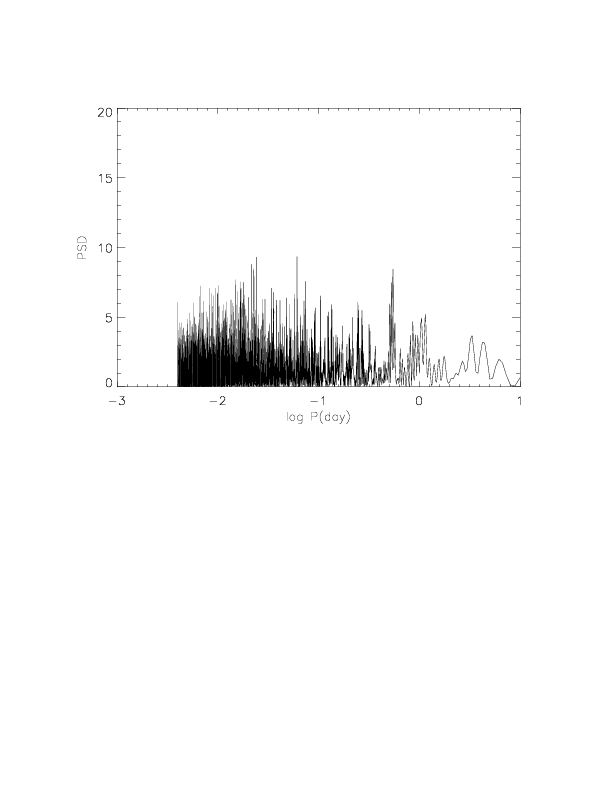}
  \caption{Periodogram from RVs for HD114174 before (left panel) and after (right panel) the subtraction of the signal at a period of 34.6~d.}
  \label{fig:rv_periodogram}
\end{figure}

\begin{figure}[hbt!]
  \centering
  \includegraphics[width=\columnwidth]{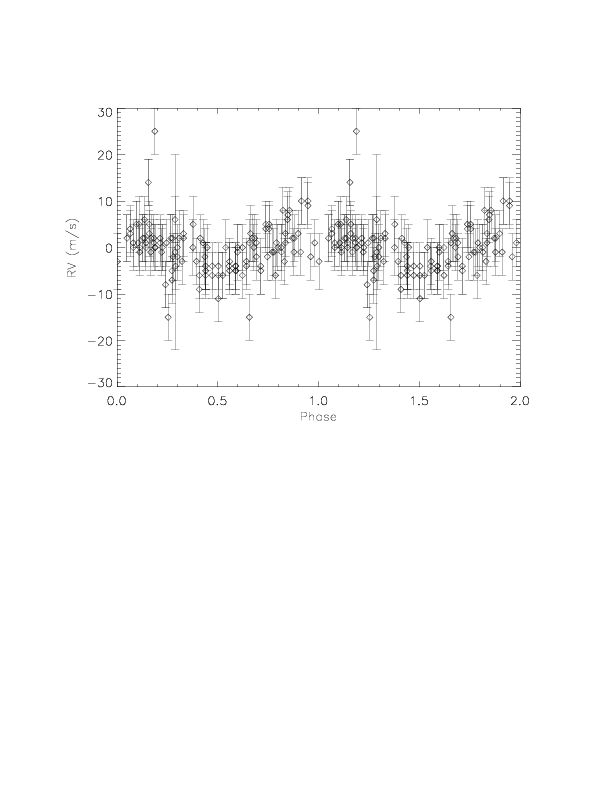}
  \caption{RV curve for HD114174 phased at a period of 34.6~d.}
  \label{fig:rv_phased}
\end{figure}

Gyrochronology relates the rotational period of late-type stars to their ages. We obtained the rotational periods for the MS stars in the various systems considered in this paper from the analysis of the TESS data \citep{tess2015}. We downloaded data from the Miskulski Archive for Space Telescopes (MAST) portal\footnote{https://archive.stsci.edu/tess/}. Here, we only considered light curves produced by the Science Processing Operations Center (SPOC: \citealt{Jenkins2016}) available on the archive on July 20, 2020. Search areas were kept at $<10$ arcsec (half a TESS pixel) around the nominal star position to avoid misinterpretation of data. The light curves are shown on the left panels of Figure~\ref{fig:tess}. Periods were determined by us from the main peak in the Scargle periodogram extracted from the light curves (right panels of Figure~\ref{fig:tess}). All stars show strong and highly significant peaks in their periodograms, though the value for HD~114174 ($\sim 40$~d) is uncertain because it is longer than the TESS time series\footnote{On the other hand, the SPOC analysis did not detect any candidate transit in these systems}. The period we obtained for CD-56~7708 (2.326 d) agrees well with that estimated by \citet{Kiraga2012} from ASAS data (2.3 d). The rich radial velocity sequence (\citealt{crepp2013, Butler2017} and HARPS data reported in the appendix) also allows one to derive the rotational period for HD~114174. Once we removed the long-term trend due to the orbital motion (see Section~\ref{sec:orbit}), we derived  the value of 34.6~d
(see Figures~\ref{fig:rv_periodogram} and \ref{fig:rv_phased}), which is in reasonable agreement with the one obtained from the TESS light curve in view of the uncertainties in this last value. We adopted the value obtained from the RV series for this star. Ages were obtained from these periods using the period-color-age calibration by \citet{Angus2019}, as modified by Bonavita et al. (2020, submitted) for stars younger than Praesepe. As discussed in Bonavita et al., these ages have errors of about 70\% (0.25 in the logarithm). Relevant data are listed in Table~\ref{tab:ages}. For comparison, \citet{Matthews2014} give a gyrochronological age of $4.0\pm 1.1$~Gyr for HD~114174, which is in quite good agreement with the present determination; however, they do not provide further details on how it was derived. As a further comparison, we would have derived ages of 17 Myr, 6.8 Gyr, and 24 Myr for HD~2133, HD~114174, and CD-56~7708, respectively, if we had used the gyro-chronological calibration by \citet{Barnes2007}, which is in quite good agreement with the values listed in Table~\ref{tab:ages}.

\subsection{WD masses, effective temperatures, and cooling ages} \label{sub:wd_fit}

\begin{figure}[hbt!]
  \centering
  \includegraphics[width=0.48\textwidth]{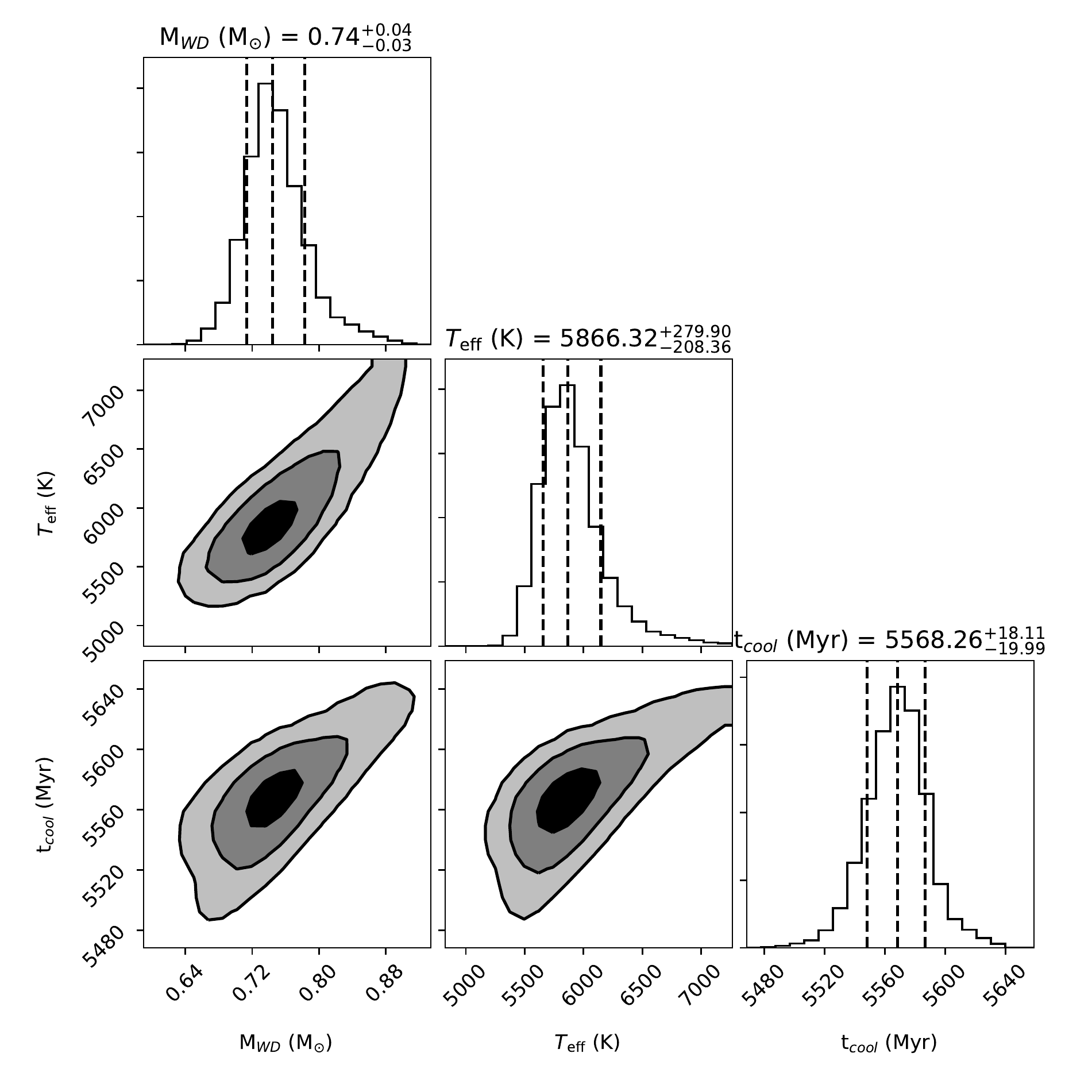}
  \caption{Marginal posterior distributions for the parameters of our model of the photometry of the WD HD114174B. We note that temperatures are in units of 1000 K.}
  \label{fig:wd_analysis}
\end{figure}

We used a Monte Carlo Markov Chain (MCMC) code to derive the main parameters of the WDs from their photometry. The routine makes use of the Python code \texttt{emcee} \citep{2013PASP..125..306F}. We ran 1,000 steps, with 20 initial walkers to constrain the three parameters of the mass (M$_{WD}$), effective temperature ($T_{\rm eff}$), and cooling time ($t_{cool}$) of each WD. The results were obtained by comparing the observed photometry with Bergeron pure hydrogen cooling sequences \citep{Bergeron1995}\footnote{www.astro.umontreal.ca/~bergeron/CoolingModels}. The best parameters obtained from the MCMC method are given in Table~\ref{tab:ages}. We note that the error bars include only the errors on the photometry, they do not include other sources of uncertainties (e.g., the parallax). They can then underestimate the real error. 

\begin{figure}[hbt!]
  \centering
  \includegraphics[width=0.48\textwidth]{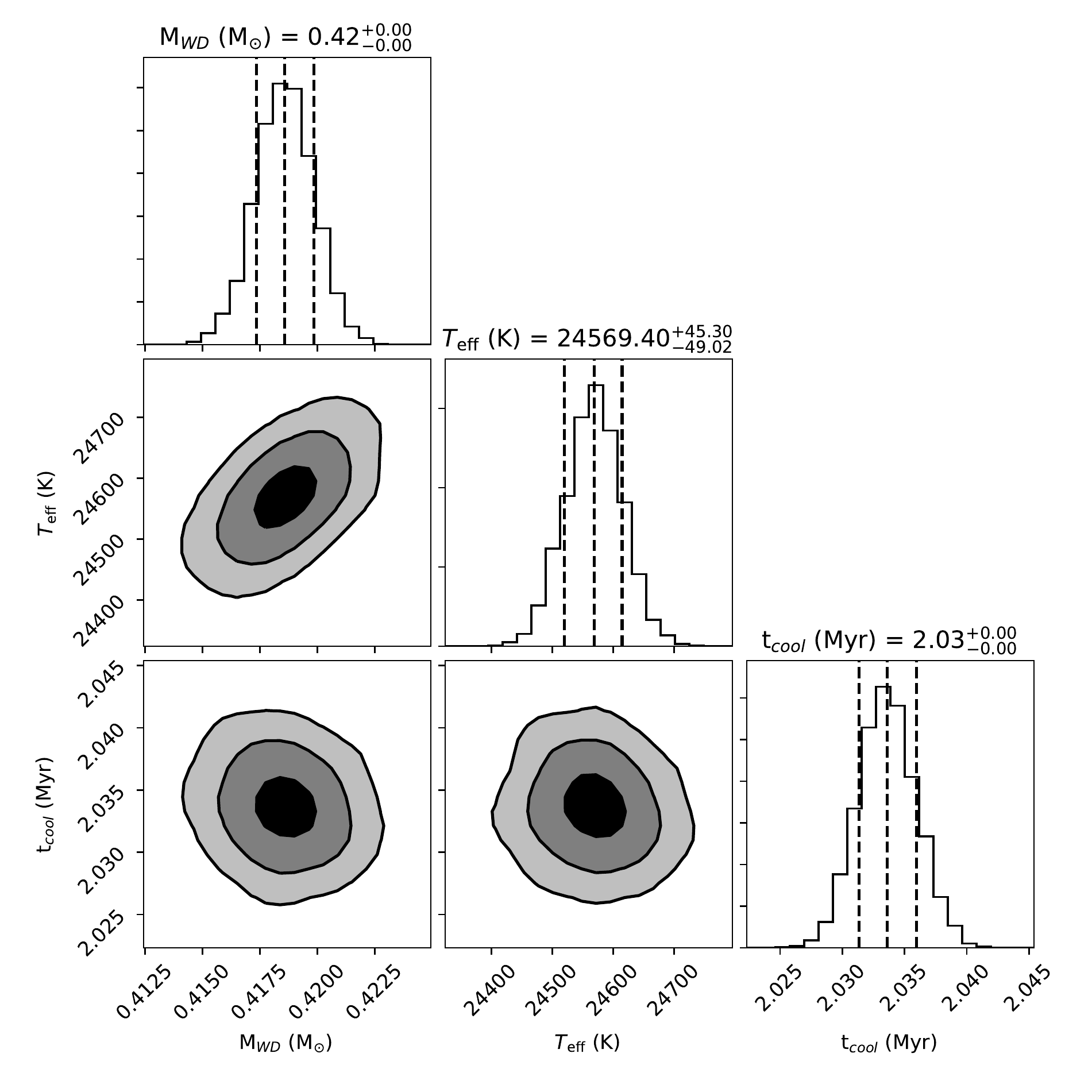}
  \caption{Marginal posterior distributions for the parameters of our model of the photometry of the WD HD2133B.}
  \label{fig:wd_analysis2}
\end{figure}

\begin{figure}[hbt!]
  \centering
  \includegraphics[width=0.48\textwidth]{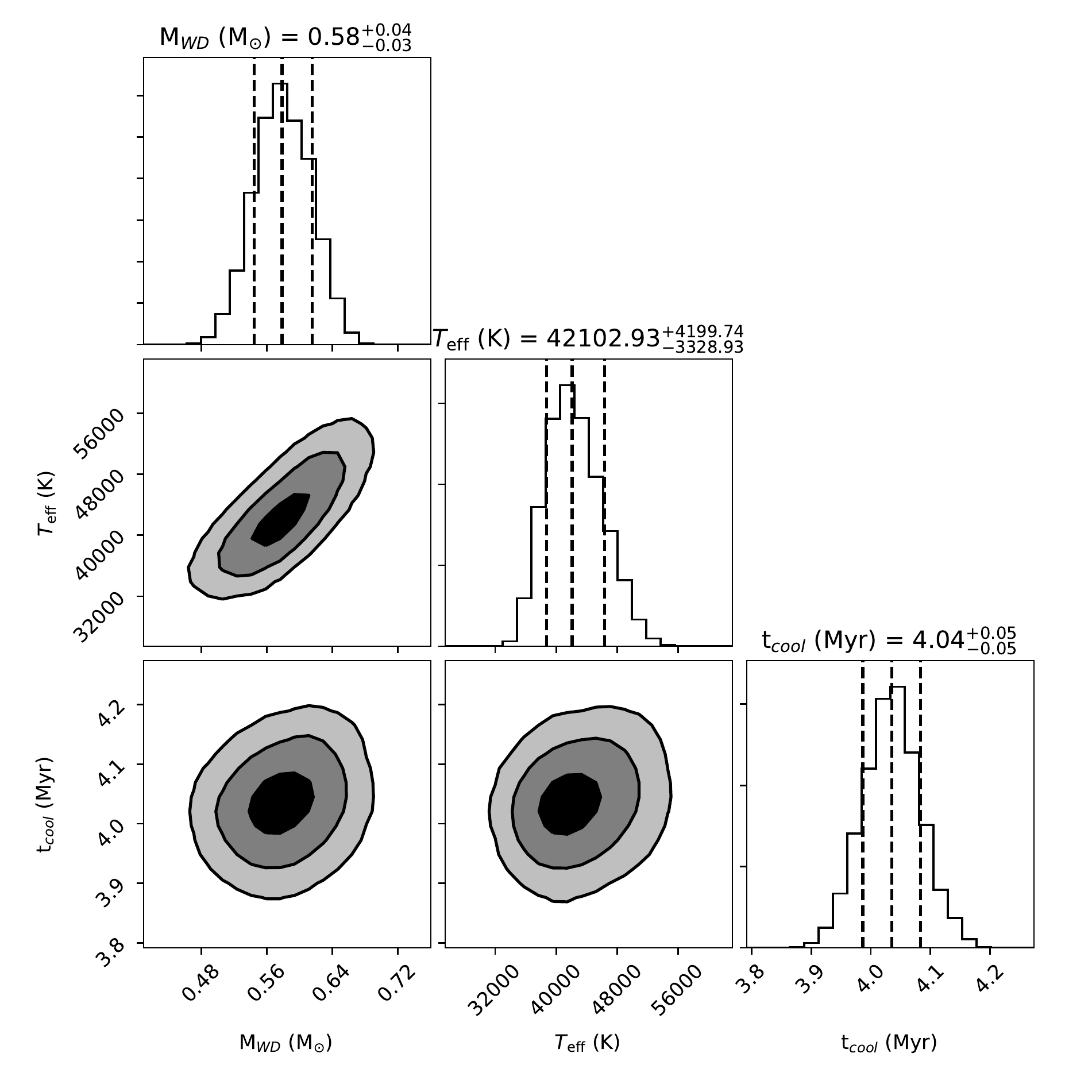}
  \caption{Marginal posterior distributions for the parameters of our model of the photometry of the WD CD-56~7708B.}
  \label{fig:wd_analysis3}
\end{figure}

Figure~\ref{fig:wd_analysis} gives the marginal posterior distributions obtained from our analysis of the photometry of HD114174B from the $y$ ($\sim 0.95~\mu$m) up to the $K_{s}$-band ($\sim 2.1~\mu$m). The extension to the $y-$band is very important because it allows one to define the effective temperature of the WD with a much higher accuracy than was possible before. In addition, the brighter magnitudes obtained in the $J$ and $H-$bands provide a rather high value of $T_{\rm eff}=5890\pm 270$~K, and consequently a smaller radius and then a higher mass ($M=0.75\pm 0.03$~$M_\odot$) for the WD. All these facts point towards a rather massive but not excessively old WD ($5.57\pm 0.02$~Gyr), which matches the remaining constraints for this system much better (see Section~\ref{sec:discussion}).

Table~\ref{tab:analogues} lists a number of WD analogs to HD~114174B \citep{Bergeron2001, Gianninas2011, Giammichele2012}; they were selected in order to be within 2$\sigma$ in $T_{\rm eff}$, mass, and age. Gaia DR2 parallaxes were used to update the absolute magnitude with respect to the catalog values, but modifications with respect the original values are small. Most of these analogs have a bluer $J-K$\ color than what we found for HD~114174B (though at only slightly more than 1-$\sigma$), and they have H-dominated spectra.

\begin{table*}[htb]
\caption{WD analogs to HD~114174B.}
\centering
\begin{tabular}{lcccccccc}
\hline
\hline
WD & $\pi$ (mas) & $M_J$ & $J-K$ & $T_{\rm eff}$ (K) & $M/M_\odot$ & Age (Gyr) & Sp. Type\\
\hline
HD~114174B & 37.91 & 13.55 & 0.44 & $5890\pm 270$ &     $0.75\pm 0.03$ & 5.57 \\
W0553+053  & 125.1 & 13.44 & 0.30 & $5785\pm 105$ & $0.72\pm 0.03$ & 4.25 & DAH \\ 
W0752-676  & 122.37 & 13.22 & 0.36 & $5735\pm 103$ & $0.73\pm 0.06$ & 4.40 & DA \\ 
W1257+037  & 60.80  & 13.48 & 0.31 & $5616\pm 100$ & $0.70\pm 0.06$ & 4.52 & DA \\ 
W1309+853  & 60.72  & 13.61 & 0.65 & $5440\pm  98$ & $0.71\pm 0.02$ & 5.45 & DAP \\ 
W1315-781  & 51.84 & 13.46 & 0.31 & $5619\pm 193$ & $0.69\pm 0.02$ & 4.39 & DC: \\ 
W1748+708  & 160.98 & 13.80 & 0.27 & $5570\pm 107$ & $0.79\pm 0.01$ & 5.86 & DXP \\ 
W2211-392  & 55.09  & 13.60 & 0.33 & $6150\pm 135$ & $0.80\pm 0.04$ & 4.26 & DA \\ 
\hline
\end{tabular}
\label{tab:analogues}
\end{table*}

Figures~\ref{fig:wd_analysis2} and \ref{fig:wd_analysis3} give the marginal posterior distributions obtained from the analysis of the fit of the photometric data, with model atmosphere and cooling sequences for the WDs in HD~2133 and CD-56~7708. Here, we consider our own as well as UV data from \citet{barstow2001} and \citet{Beitia-Antero2016}. For comparison, \citet{Joyce2018} used FUSE spectra and HST data to derive the mass, radius, and temperatures for the WDs in both systems. They obtained temperatures of $T_{\rm eff}=29724\pm 158$\ and $49037\pm 263$~K for HD~2133B and CD-56~7708B, respectively, which is a bit warmer than the values derived by \citet{holberg2013} and in our analysis. The mass of HD~2133B was not well determined in the study by \citet{Joyce2018}, with a very low mass obtained from FUSE data ($0.28\pm 0.07$~$M_\odot$) and a value of $0.40\pm 0.14$~$M_\odot$, which is more consistent with expectations and our analysis from HST data. Their mass for CD-56~7708B ($0.524\pm 0.04$~$M_\odot$) is a bit smaller but within the error bars of our determination. The cooling age for CD-56~7708B agrees fairly well with the estimate by \citet{Barstow2014} of 1.6 Myr. The cooling ages for both HD~2133B and CD-56~7708B are both shorter than those estimated from gyrochronology. However, uncertainties in gyrochronological ages at very young ages are large. On the other hand, there is very good agreement between the cooling age of the WD and the gyrochronology age of the MS star for HD114174. Anyhow, we should take into account that the assumption of re-rejuvenated MS stars in Sirius-like systems being very similar to very young stars is possibly inaccurate.

\section{Spectroscopic properties of the primary stars} \label{sec:primaries}

The atmospheric parameters of the host stars used for the spectroscopic analysis were computed using photometric calibrations. The $V$ band magnitudes were obtained  by a transformation (J. Yana Galarza, private communication) that uses $G$ band photometry from Gaia \citep{Gaia2016}, resulting in $V$ magnitudes that are more precise than previously published values. The distances were derived from the Gaia parallax, and by using the galactic latitude and longitude, we estimated the color excess from Stilism (STructuring by Inversion the Local Interstellar Medium) maps \citep{Lallement2014}. 

We used the $V$ and 2MASS $JHK$ photometry \citep{Skrutskie2006} and the $E(B-V)$ obtained above and extinction ratios from \cite{Ramirez2005} to compute the intrinsic colors $(V-J)$, $(V-H),$ and $(V-Ks)$. Then, we estimated the effective temperature using the color-metallicity-$T_{\rm eff}$ calibrations by \cite{Casagrande2010}.

In addition to the mass, $T_{\rm eff}$, parallax, and $V$ magnitude, the bolometric corrections ($BC_V$) from \cite{Masana2006} were considered to determine the trigonometric surface gravity ($\log{g}$) of the primary star. Finally, the micro-turbulence ($v_{\rm t}$) and macro-turbulence ($v_{\rm macro}$) were determined using the relations by \cite{Ramirez2014} and \cite{dosSantos2016}, respectively. The 2017 version of the LTE code MOOG \citep{sneden1974} was used to determine the metallicity, using Kurucz model atmospheres \citep{castelli2004new}. As the above calibration relations depend on metallicity, the stellar parameters listed in Table~\ref{tab:finalAtm} were determined iteratively.

We also tried the spectroscopic equilibrium method \citep[e.g.,][]{Melendez2012}, but in the two highly-rotating stars, HD 2133 and CD-56 7708, many important lines could not be measured due to blending caused by the badly broadened profiles, making the method unfeasible. For the star with low rotation (HD 114174), the spectroscopic result is compatible with the method applied above based on photometric temperatures and trigonometric gravities.

\begin{table*}[htb]
\centering
\caption{Adopted parameters and abundances for the MS stars.}
\begin{tabular}{lccc}
\hline \hline
 & HD 114174 & HD 2133 & CD-56 7708\\ \hline
 $T_{\rm eff}$ (K) & $5703 \pm 31$ & $5991 \pm 43$ & $ 5545 \pm 42$\\ 
 $BC_V$ (mag)& $-0.13 \pm 0.01$ & $ -0.07 \pm 0.01 $ & $-0.15 \pm 0.02$\\ 
$\log{g}$ (cm s$^{-2}$) & $ 4.35\pm 0.02 $ & $ 4.28 \pm 0.04 $ & $4.42 \pm 0.06$\\ 
$v_{\rm t}$ (km s$^{-1}$) & $0.97 \pm 0.12 $ & $ 1.22 \pm 0.12 $ & $0.79 \pm 0.12$\\ 
$v_{\rm macro}$ (km s$^{-1}$) & $3.16 \pm 0.06 $ & $ 4.36 \pm 0.10 $ & $2.50 \pm 0.04$\\
$v \sin{i}$ (km s$^{-1}$) & $1.71 \pm  0.12$ & $24.37 \pm 0.93$ & $24.9 \pm 0.06$\\
 $[$Fe/H$]$ (dex) & $0.016 \pm 0.037$ & $ 0.080 \pm 0.062$ & $ 0.380 \pm 0.064 $\\ 
  $[$C/H$]$ (dex) & $0.030 \pm 0.065$ & $-0.050 \pm 0.149$ & $ 0.24 \pm 0.15$ \\ 
$[$Y/H$]$ (dex) & $0.370 \pm 0.108$ & $ 0.100 \pm 0.127$ & $ 1.26 \pm 0.13$ \\ 
$[$Ba/H$]$ (dex) & $0.220 \pm 0.087$ & $ 0.190 \pm 0.093$ & $ 0.91 \pm 0.12$ \\ 
\hline
\end{tabular}
\label{tab:finalAtm}
\end{table*}

Chemical abundances were determined using spectral synthesis by comparison between observed and theoretical spectra. The atmosphere models were interpolated with the $q^2$ code \citep{Ramirez2014} and by using Kurucz model atmospheres \citep{castelli2004new}. The line profiles were analyzed using the 2017 version of MOOG \citep{sneden1974}, using the line list of \citet{Melendez2012}. 

The abundances from the solar analogs stars were compared to the solar abundances, using a line-by-line differential analysis method \citep{Gratton2001,Bedell2014,Biazzo2015}, but using spectral synthesis rather than equivalent widths. In order to determine the metallicity, we chose ten Fe~I absorption lines (4950.1, 5373.7, 5679.0, 5934.7, 6003.0, 6065.5, 6173.3, 6252.6, 6265.1, and 6430.8 $\mbox{\AA}$). 
The criterion was to get relatively clean lines (mostly unblended) in the three solar analogs, meaning that preferentially strong lines were selected, as the weak lines are washed out in the spectra of HD 2133 and CD-56 7708 due to their fast rotation ($\sim 25$ km $s^{-1}$). 
The $v$ sin $i$ was determined by spectral synthesis of the ten iron lines above, including the instrumental broadening (FWHM from the resolving power of each spectrograph) and $v_{\rm macro}$.
HD~114174 is a solar twin \citep{Ramirez2014}  with [Fe/H] $= 0.016 \pm 0.037$, HD~2133 is a solar analog with [Fe/H] $= 0.080 \pm 0.062,$ and CD-56 7708 is a high metallicity star with [Fe/H] $= 0.380 \pm 0.064$. 

\begin{figure}[hbt!]
  \centering
  \begin{tabular}{cc}
  \includegraphics[width=0.22\textwidth]{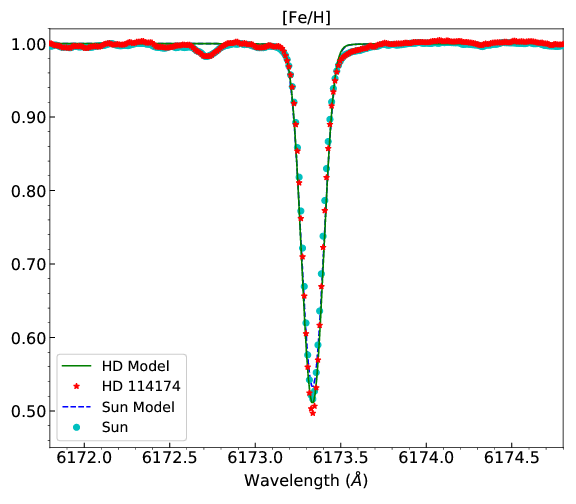}   & 
  \includegraphics[width=0.22\textwidth]{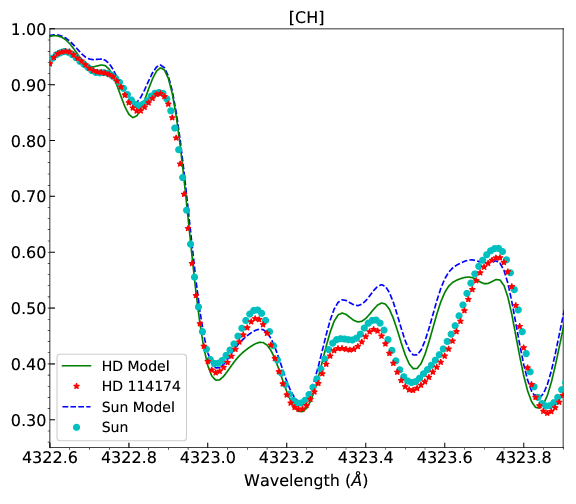}  \\
  \includegraphics[width=0.22\textwidth]{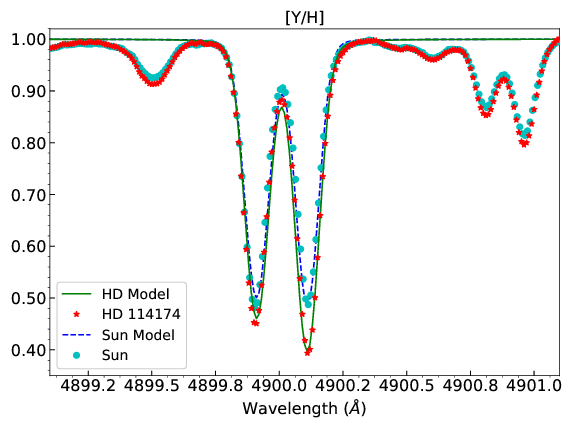}   &
  \includegraphics[width=0.22\textwidth]{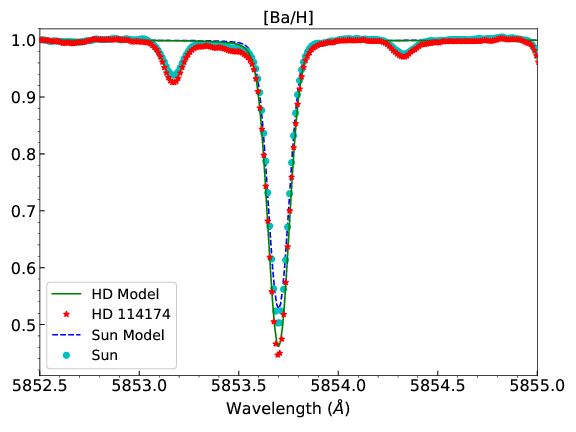} \\
  \end{tabular}
  \caption{Spectral synthesis of features used to derive the values of [Fe/H], [C/H], [Y/H], and [Ba/H] for HD~114174, and a comparison with the solar spectrum.}
  \label{fig:sint_hd114}
\end{figure}

\begin{figure}[hbt!]
  \centering
  \begin{tabular}{cc}
  \includegraphics[width=0.22\textwidth]{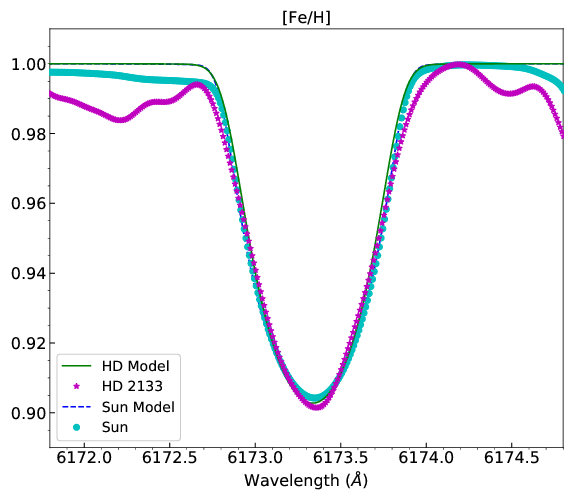} &
  \includegraphics[width=0.22\textwidth]{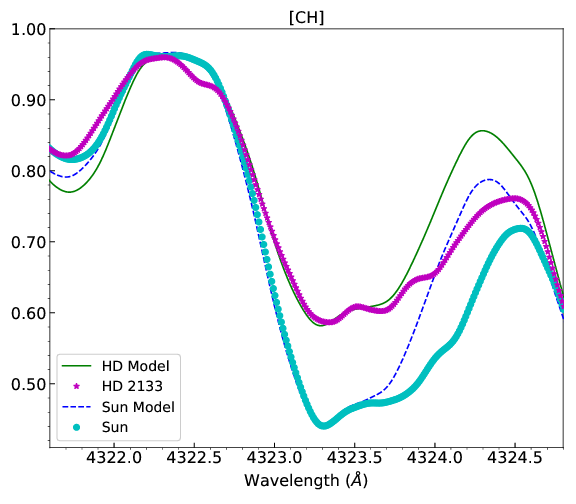} \\
  \includegraphics[width=0.22\textwidth]{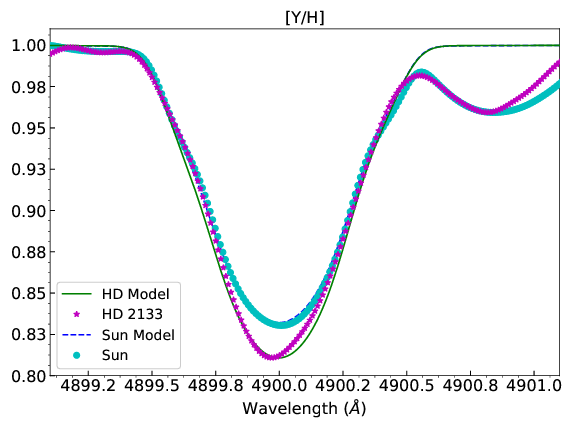} &
  \includegraphics[width=0.22\textwidth]{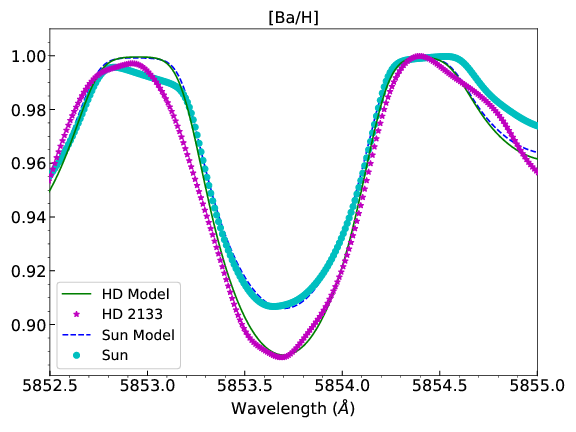} \\
  \end{tabular}
  \caption{Spectral synthesis of features used to derive the values of [Fe/H], [C/H], [Y/H], and [Ba/H] for HD 2133, and a comparison with the solar spectrum broadened to FWHM = 25 km s$^{-1}$. It is important to notice that the x-scale is the same as in Fig. \ref{fig:sint_hd114}.}
  \label{fig:sint_hd213}
\end{figure}

\begin{figure}[hbt!]
  \centering
  \begin{tabular}{cc}
  \includegraphics[width=0.22\textwidth]{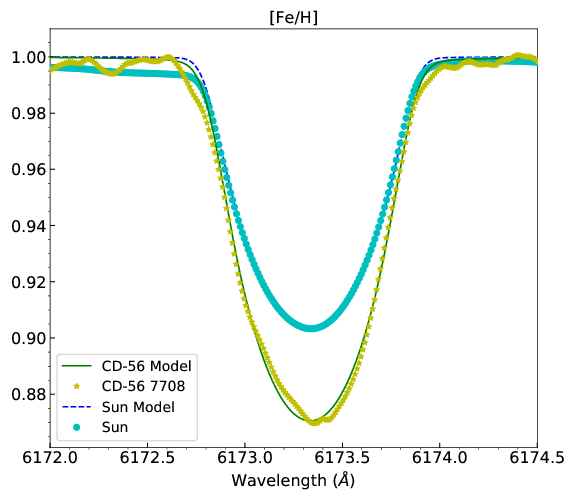}   &
  \includegraphics[width=0.22\textwidth]{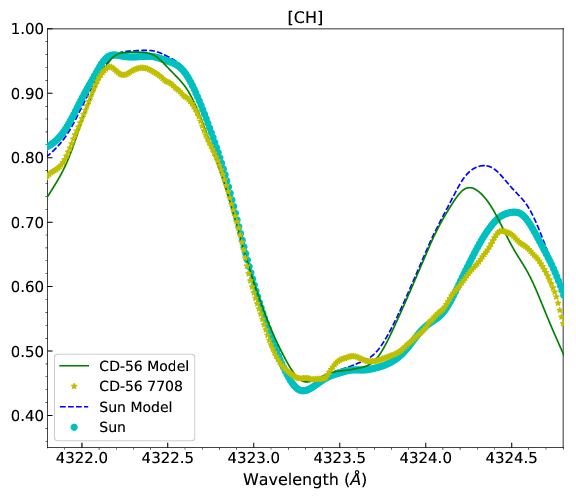} \\
  \includegraphics[width=0.22\textwidth]{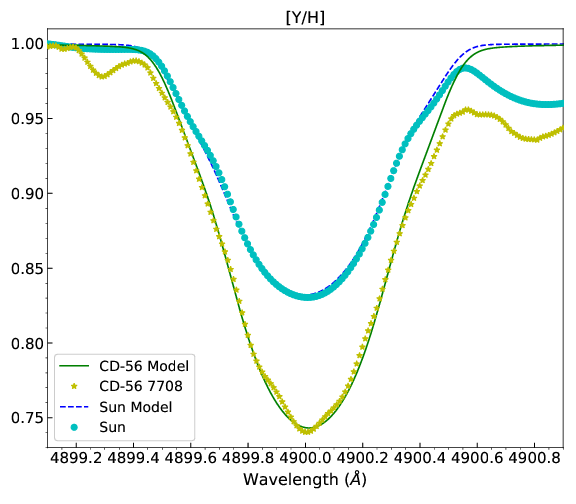}   &
  \includegraphics[width=0.22\textwidth]{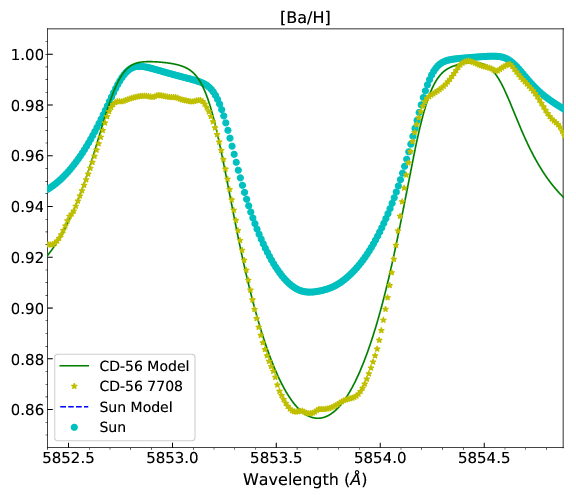} \\
  \end{tabular}
  \caption{Spectral synthesis of features used to derive the values of [Fe/H], [C/H], [Y/H], and [Ba/H] for CD-56~7708, and a comparison with the solar spectrum broadened to FWHM = 25 km s$^{-1}$. It is important to notice that the x-scale is the same as in Fig. \ref{fig:sint_hd114}.}
  \label{fig:sint_cd567}
\end{figure}

Using the magnitude difference between the primary and the secondary stars, we estimated the contamination fraction of the WD on the MS star. Only the system with the coolest MS and hottest WD, CD-56~7708, has a high contamination by the WD. For the V band, this was estimated using the observed magnitudes and extrapolated toward the blue using the expected colors for objects of the same temperatures as the MS star and the WD. The contamination results of $\sim 6\%$ in the blue region and is progressively lower towards redder wavelengths ($\sim 2\%$ in the $V-$band). For this star, we took the contamination by the WD companion into account when determining the chemical abundances.

The abundances of C and neutron capture elements are important when verifying whether the stars have excesses of these elements and to get a better constraint in the comparison with yields of AGB models. The C atomic lines available for our spectra are not detectable on the rapidly rotating stars, so we chose a molecular region of CH around 4323.2 $\mbox{\AA}$ to measure the C abundance. As far as what concerns n-capture elements, we chose the Y line at 4900.1 $\mbox{\AA}$ and the Ba line at 5853.7 $\mbox{\AA}$, considering the hyperfine structure and blends. The resulting [C/H], [Y/H], and [Ba/H] ratios are shown in Table~\ref{tab:finalAtm}. It is important to notice that in order to determine a better differential abundance for the two highly-rotating stars, the solar spectrum used for the reference solar abundances was broadened with a FWHM = 25 km~s$^{-1}$.

\begin{figure}[hbt!]
  \centering
  \begin{tabular}{cc}
  \includegraphics[width=0.22\textwidth]{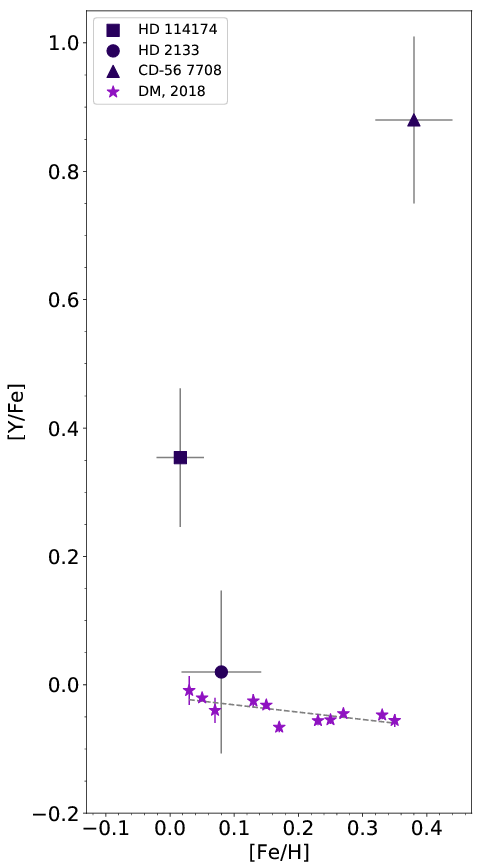} &
  \includegraphics[width=0.22\textwidth]{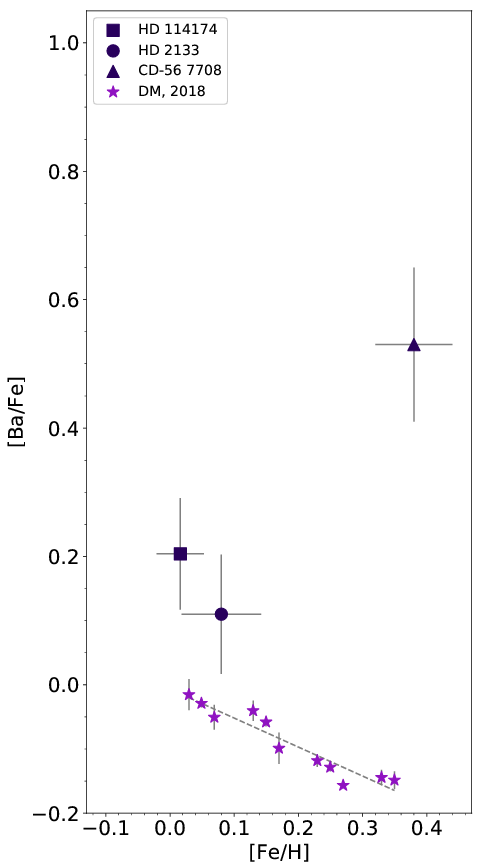} \\
  \end{tabular}
  \caption{Y (left) and Ba (right) abundances as a function of metallicity [Fe/H] for our three Solar-type stars with WD companions (black symbols with error bars), compared to abundances for normal Solar-type stars determined by  \cite{Mena2018}.}
  \label{fig:over_abund}
\end{figure}

We compared the Y and Ba abundances derived for HD~114174, HD~2133, and CD-56~7708 to the abundances derived by \cite{Mena2018} for Solar-type stars. In Figure~\ref{fig:over_abund}, we can see a clear overabundance for our Solar-type stars with WD companions, relative to other Solar-type stars that follow the standard chemical evolution of the Galaxy; in particular, CD-56~7708 is clearly a mild (dwarf) Ba-star, while excesses for the two other stars are small.

Finally, we checked all of our spectra for the presence of the Li~{\sc i} doublet at 6707.78 \AA. As expected, this feature is undetected in all our target stars.

\section{Orbital study} \label{sec:orbit}

\subsection{HD~114174}

\begin{figure*}[hbt!]
  \centering
  \includegraphics[width=\textwidth]{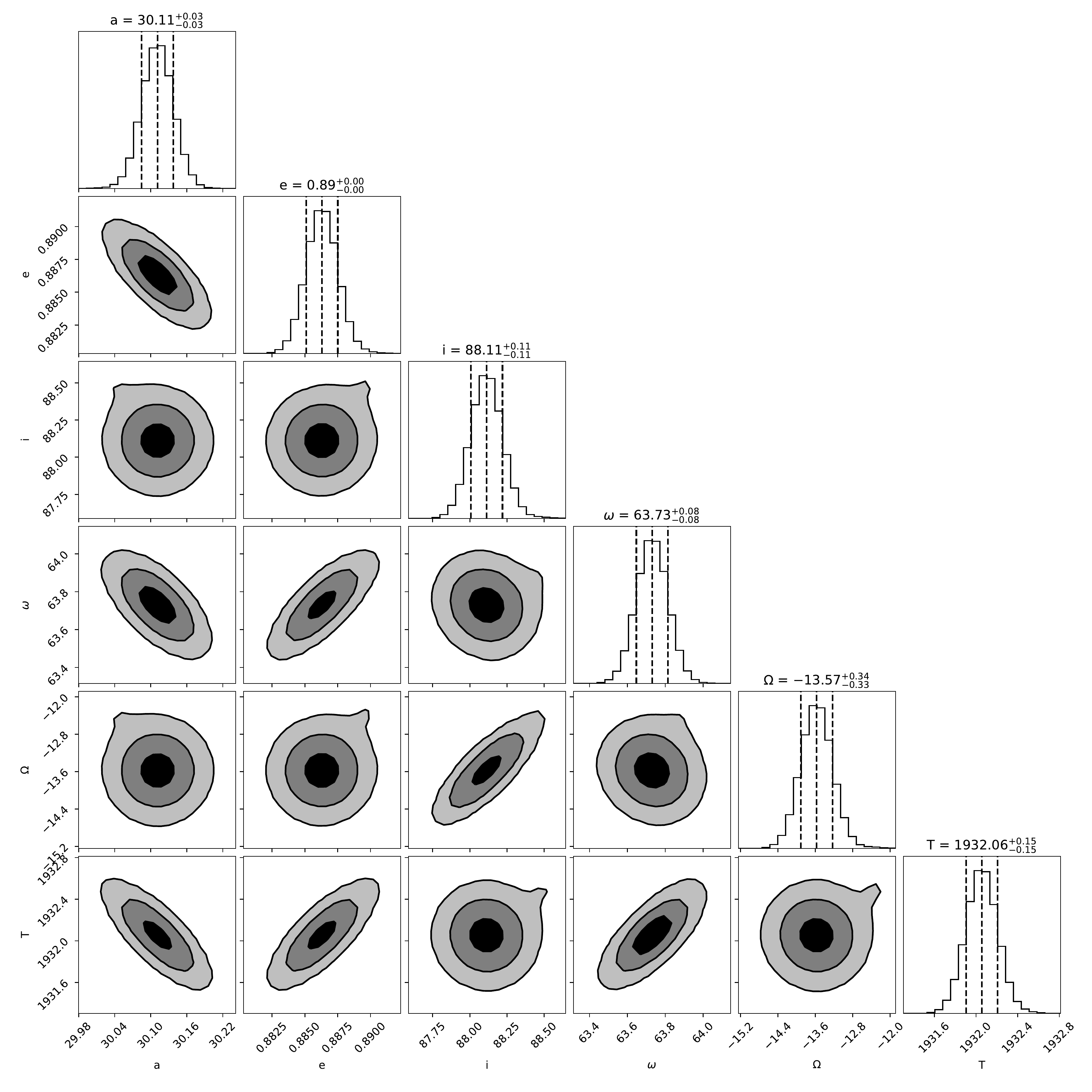}  
  \caption{Marginal posterior distributions for the parameters of our model of the orbit of HD114174 obtained with the MCMC method.}
  \label{fig:post_HD114174}
\end{figure*}

\begin{table}[htb]
\centering
\caption{Astrometry for HD~114174 used in this paper.}
\begin{tabular}{lcccc}
\hline \hline
JD       &   PA   &  Sep  & err   & ref \\
         & degree &  arcsec & arcsec & \\
\hline
55615.1  &  171.8  & 0.7198 & 0.0066 & 1 \\
55960.1  &  172.1  & 0.7011 & 0.0050 & 1 \\
56076.8  &  170.5  & 0.6958 & 0.0058 & 1 \\ 
56112.8  &  172.2  & 0.6921 & 0.0087 & 1 \\
56792.1  &  172.   & 0.6547 & 0.0020 & \\
58198.69 &  174.   & 0.5770 & 0.0300 & 2 \\
57058.37 &  172.03 & 0.6405 & 0.0020 & 2 \\
57112.20 &  171.71 & 0.6372 & 0.0020 & 2 \\
57146.09 &  172.02 & 0.6356 & 0.0020 & 2 \\
57174.05 &  172.24 & 0.6341 & 0.0020 & 2 \\
57180.12 &  171.60 & 0.6356 & 0.0020 & 2 \\
57207.95 &  172.00 & 0.6348 & 0.0020 & 2 \\
57208.04 &  172.15 & 0.6355 & 0.0020 & 2 \\
57792.22 &  172.27 & 0.6001 & 0.0020 & 2 \\
57831.83 &  172.30 & 0.6044 & 0.0020 & 2 \\
57882.15 &  171.88 & 0.5911 & 0.0020 & 2 \\
58220.08 &  172.42 & 0.5773 & 0.0020 & 2 \\
58244.01 &  172.56 & 0.5739 & 0.0020 & 2 \\
58255.96 &  172.42 & 0.5740 & 0.0020 & 2 \\
58549.15 &  172.34 & 0.5594 & 0.0020 & 2 \\
58553.13 &  172.64 & 0.5589 & 0.0020 & 2 \\
58587.08 &  172.61 & 0.5536 & 0.0020 & 2 \\
58621.99 &  172.58 & 0.5535 & 0.0020 & 2 \\
\hline
\end{tabular}
\label{tab:astro_hd114714}

References: 1: \citet{crepp2013}; 2: This paper (observation on 58198.69 is with GPI/Gemini under proposal ID GS-2018A-FT-103, PI Pacheco; all other observations are with SPHERE at VLT).
\end{table}

\begin{table}[htb]
\centering
\caption{Orbit for HD~114174.}
\begin{tabular}{lcc}
\hline \hline
Parameter       & {\it Orbit}   &  MCMC   \\          
\hline
a (au)          &  28.5$\pm$40  &  30.11$\pm$0.03 \\   
P (yr)          &  76.8$\pm$9.2 &       124       \\   
T0 (yr)         & 1974.6$\pm$6.4  & 1932.06$\pm$0.25\\   %
e               & 0.91$\pm$32  &   0.89$\pm$0.00 \\   
$\Omega$ (degree) &-12.0$\pm$1.5 & -13.57$\pm$0.35 \\ 
$\omega$ (degree) &  71$\pm$28  &  63.73$\pm$0.08 \\ 
i (degree)      & 88.7$\pm$2.0&  88.11$\pm$0.11 \\   
\hline
\end{tabular}
\label{tab:orb_hd114714}
\end{table}

Data for HD~114174 are quite rich, since not only do we have a rather long astrometric series (see Table~\ref{tab:astro_hd114714}), but also 131 high precision radial velocities (\citealt{crepp2013, Butler2017}, and those that can be obtained from the archive HARPS data that are reported in the appendix). Combining the data sets of both accurate astrometry and radial velocities allows for the determination of a preliminary orbit with reasonable errors. For this purpose, we used two different methods. The first one is based on the {\it Orbit} fitting code by \citet{Tokovinin2016}\footnote{https://zenodo.org/record/61119}, which is based on a Levenberg-Marquard optimization algorithm to find the best astrometric orbit. The second one is based on an MCMC analysis using a code that combines astrometric and radial velocity data at the same time. The code makes use of \texttt{emcee}. In this second case, we adopted masses of $M_{\rm MS} = 0.98$~$M_\odot$ and $M_{\rm WD} = 0.75$~$M_\odot$, as given by the analysis of the position of the MS star in the color-magnitude diagram and of the photometry of the WD (see Section~\ref{sec:ages}). Here, we assumed an error bar of 5 m/s for the RVs to take the jitter due to stellar activity  into account.

\begin{figure}[hbt!]
  \centering
  \includegraphics[width=8.8truecm]{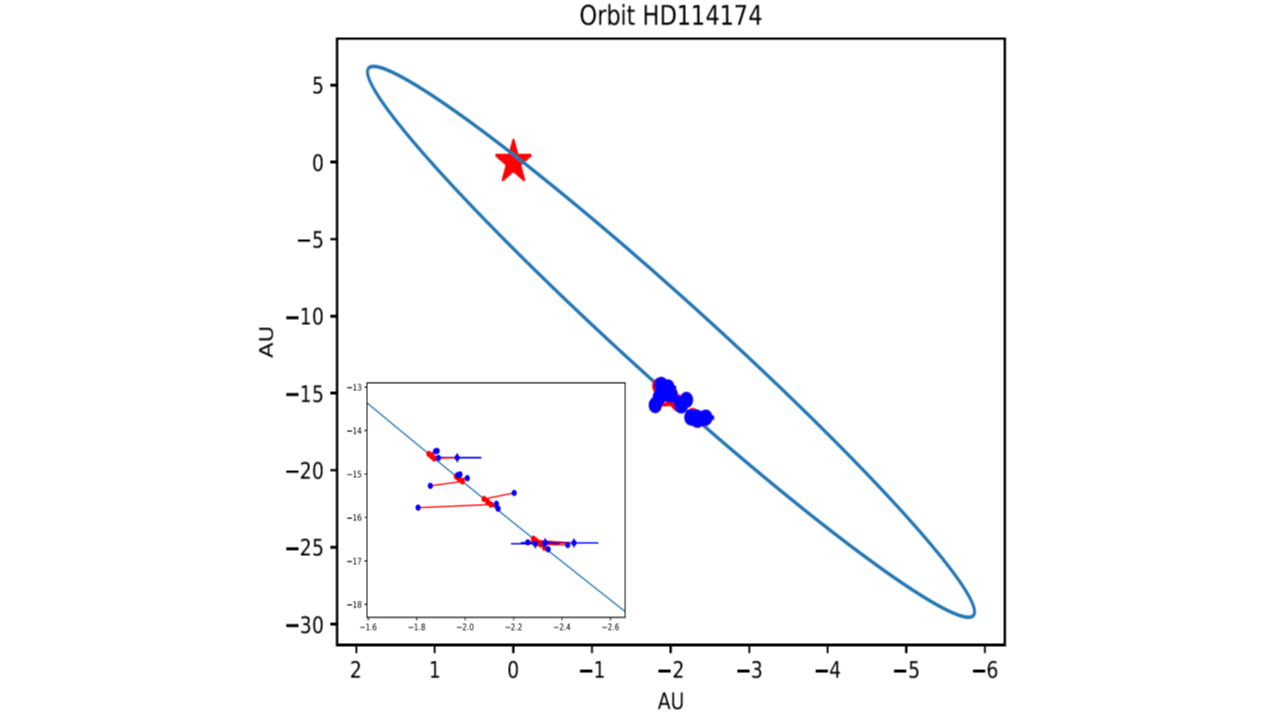}  
  \includegraphics[width=8.8truecm]{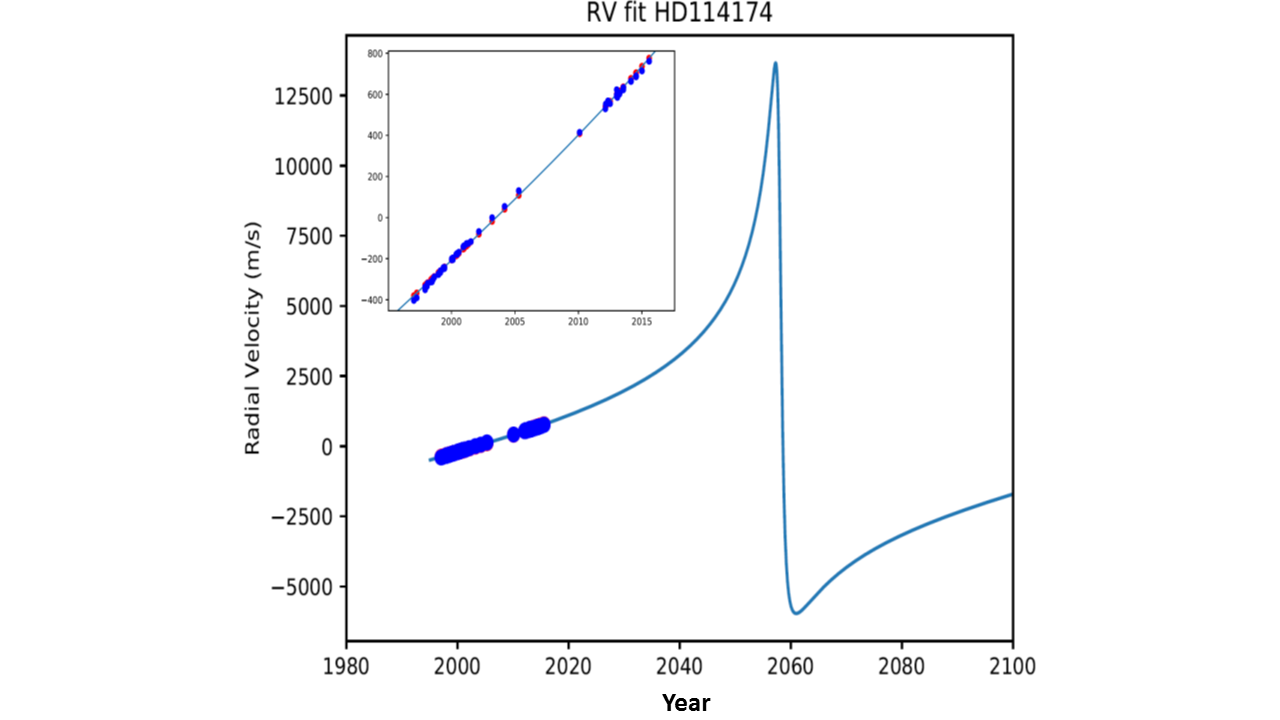}  
  \caption{Upper panel: Best orbit of HD114174 obtained with the MCMC method on the sky plane. We note that scales along RA and declination are very different; we display the orbit in this way to avoid too much confusion on the horizontal axis of the plot. Bottom panel: Radial velocity orbit. Data are shown in blue, and the model is in red. Insets in both panels show zooms of the region of the orbit covered by observations}
  \label{fig:orbit_HD114174}
\end{figure}

The results obtained with the two codes are quite similar, albeit the methods used are completely different. In Figure~\ref{fig:post_HD114174} we show the posterior distribution obtained with the MCMC method; this figures shows that once the mass of the components are assumed, the solution is quite robust despite the small section of the orbit covered; this is because of the high precision of the measures (see Figure~\ref{fig:orbit_HD114174}). In particular, it is well established that the orbit is quite eccentric and seen almost edge-on.

These orbits agree fairly well with those considered by \citet{bacchus2017}, who however made different assumptions about the MS and WD masses. All of these orbits share a high eccentricity and a very high inclination, which are quite close to edge-on.

One of the purposes for determining the orbit is to be able to determine the mass of the WD, which is to be compared with that derived from evolutionary models. In this case, we can only use the result provided by the {\it orbit} code because stellar masses were adopted to allow for the convergence of the MCMC procedure. The nominal solution found gives a total mass of the system of 3.9~$M_\odot$ with a very large uncertainty, so that this result cannot be considered significant. 

\subsection{HD~2133 and CD-56 7708}

\begin{figure}[hbt!]
  \centering
  \begin{tabular}{cc}
  \includegraphics[width=4.25truecm]{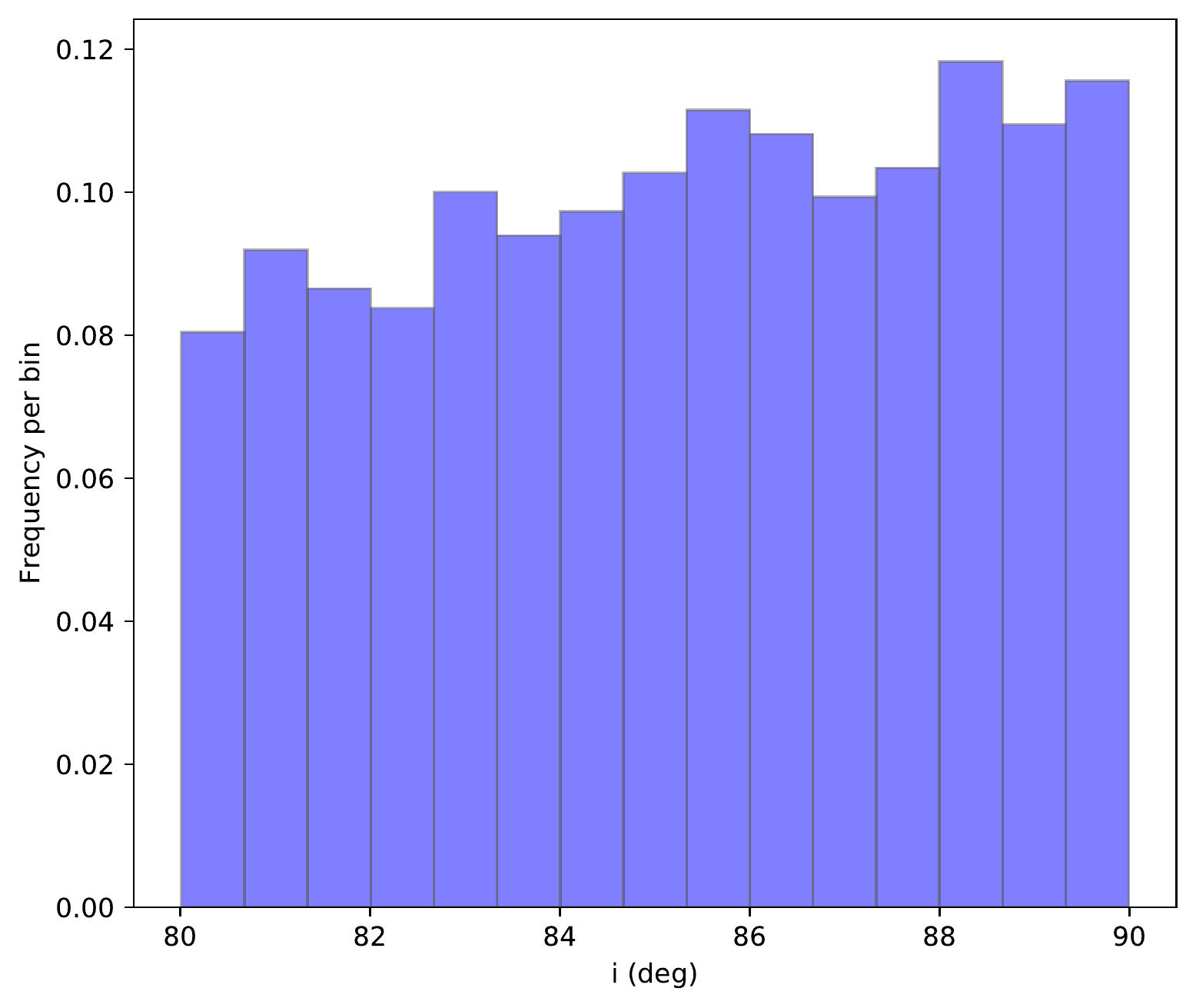} &
  \includegraphics[width=4.25truecm]{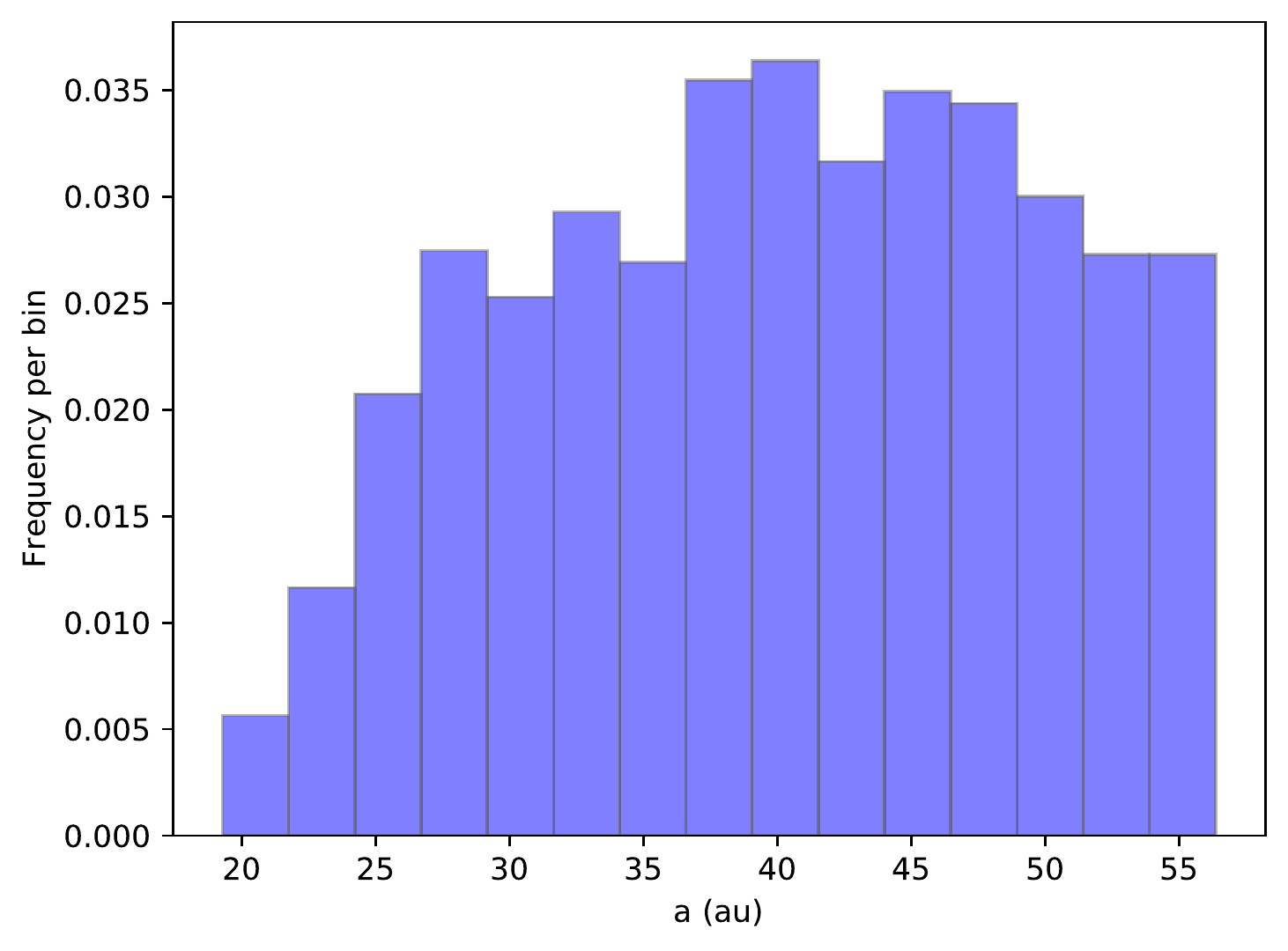} \\
  \includegraphics[width=4.25truecm]{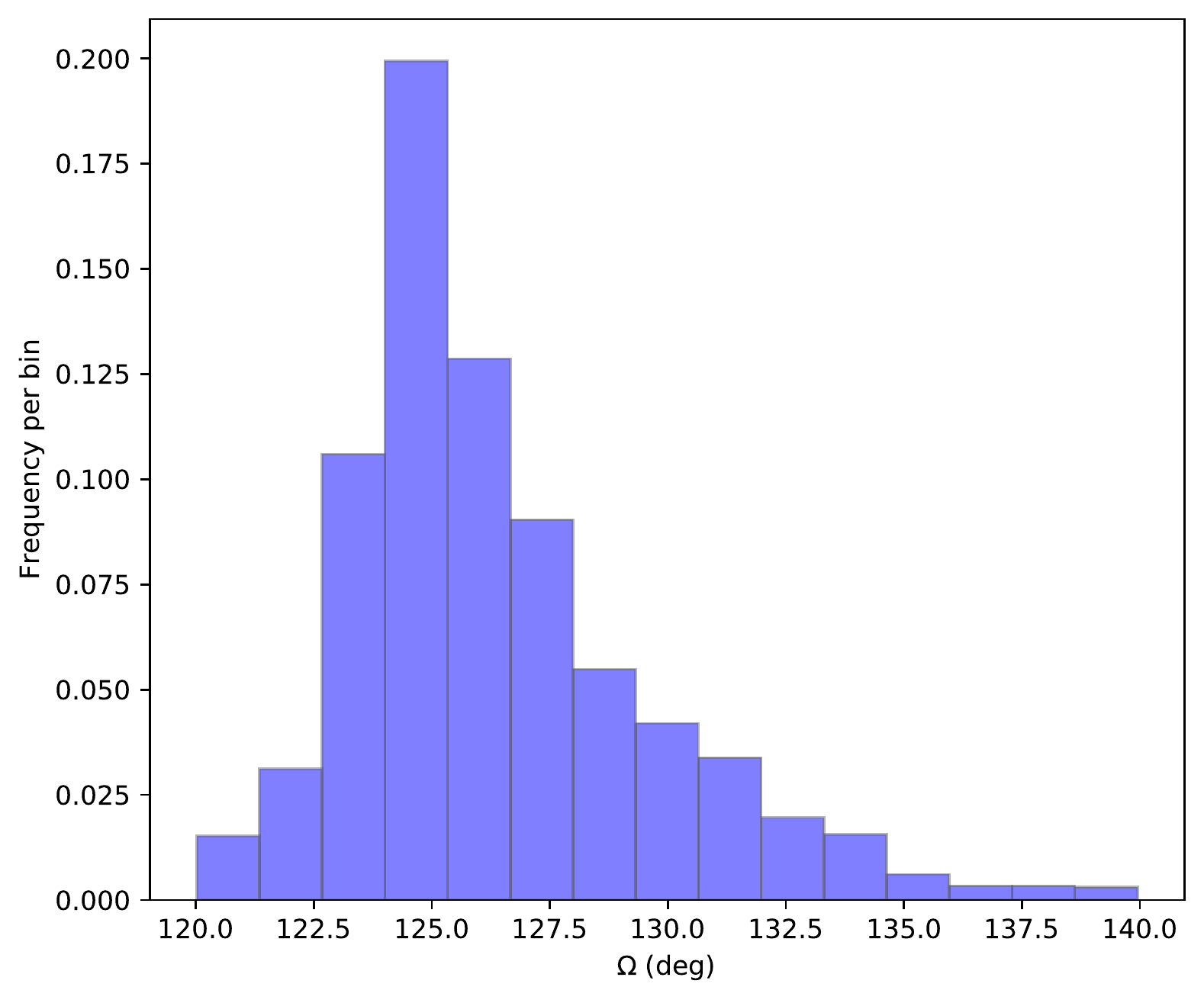} &
  \includegraphics[width=4.25truecm]{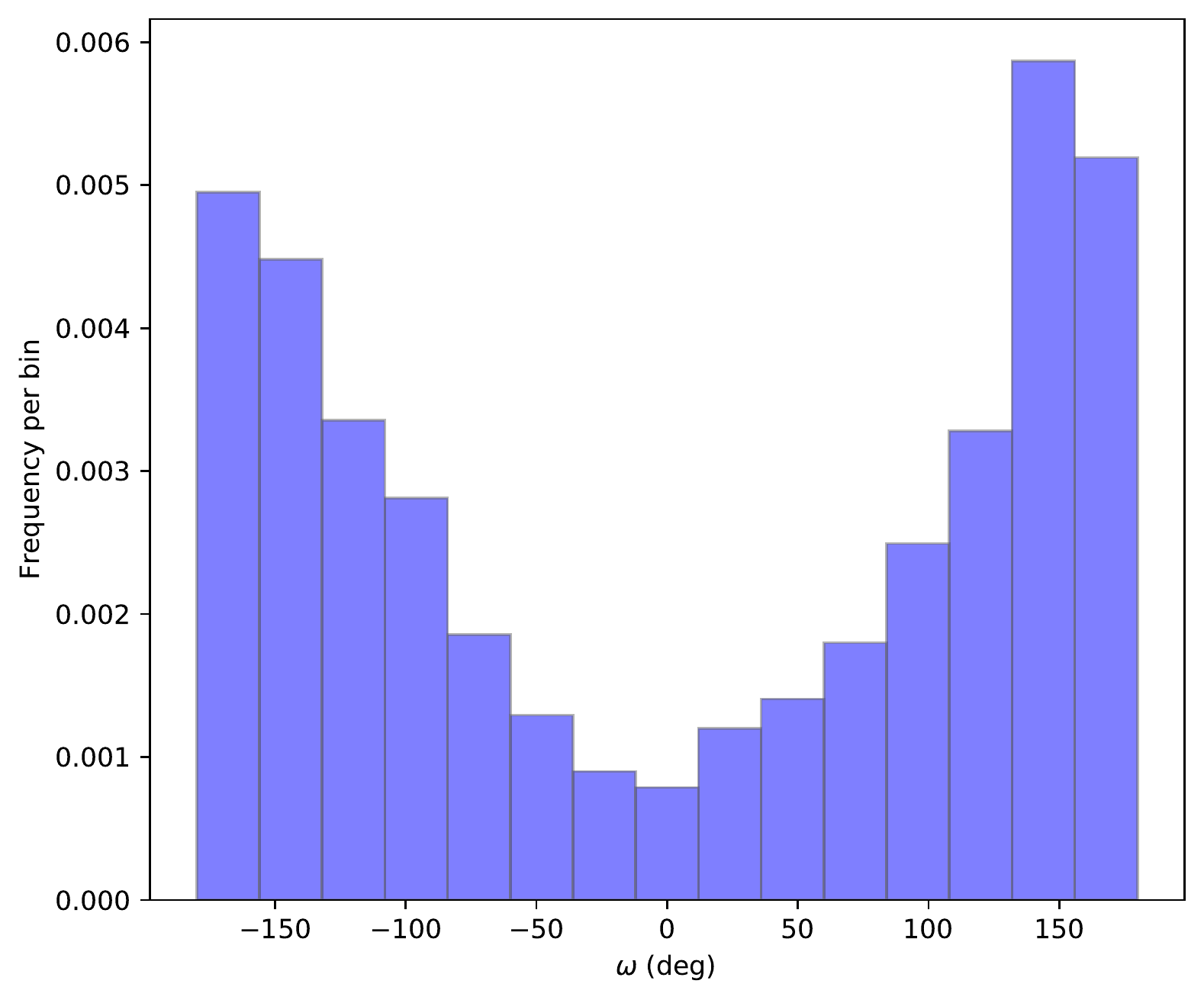} \\
  \end{tabular}
  \caption{Histograms of the frequency distribution of possible values of orbital parameters obtained with our Monte Carlo code for CD-56~7708. Upper left panel: Orbital inclination. Upper right: Semi-major axis.\ Lower left: The position angle of the ascending node $\Omega$. Lower right: The argument of periastron $\omega$.}
  \label{fig:orbit_CD-567708}
\end{figure}

Much less data are available for HD~2133 and CD-56 7708; these stars are much further from the Sun, making astrometric motion more difficult to constrain even extending the series considering the HST data by \citet{barstow2001}, and there are not adequate high precision RV series. Lacking data for a full orbit determination, \citet{holberg2013} proposed tentative periods of 665 and 118 yr for HD~2133 and CD-56 7708, respectively, from a statistical analysis, where eccentricity, inclination, and other parameters have specific distributions. These values should be considered as order of magnitude estimates, at best.

To estimate the probability distribution of the orbital parameters, we ran a Monte Carlo simulation using the code presented in \cite{Zurlo2018}. This code explores different orbits that can fit the astrometric data when a small fraction of the orbit is covered by the observations. Using this code, we could not obtain any significant constrain for HD~2133. Something more can be said for CD-56~7708; when assuming masses of $M_{\rm MS}=0.93$ and $M_{\rm WD}=0.58$~$M_\odot$ for the two components (see Section~\ref{sec:ages}), we obtained the distributions of possible  orbital parameters shown in Figure~\ref{fig:orbit_CD-567708}. We found that the semimajor axis $a$\ is in the range of 20 - 55 au, the orbital inclination $i$ is $>80$~degrees (i.e., the orbit is seen quite close to edge-on), the position angle of the ascending node $\Omega$ is in the range of 120 - 135 degrees, and the argument of periastron $\omega$ is more likely to be at values far from zero. On the other hand, we did not obtain any significant constraint on the orbital eccentricity.

\section{Discussion and concluding remarks} \label{sec:discussion}

\subsection{WD progenitors}

The system for which we can make more inferences is HD~114174. Using the mass we derived for HD~114174B ($0.75\pm 0.05$~$M_\odot$) and the initial-final mass relation \citep{El-Badry2018, Cummings2018}, we infer that the MS progenitor of this WD had a mass of $3.2\pm 0.4$~$M_\odot$. According to the PARSEC evolutionary models \citep{Bressan2012}, the pre-WD evolutionary time of a solar metallicity star (as determined by our abundance analysis) with this mass is $0.46\pm 0.10$~Gyr. The error bar is given by the uncertainty in the mass. The total age of the system is obtained by summing this age to the WD cooling time, which according to our determination is $5.57\pm 0.02$~Gyr. The total age is then $6.03\pm 0.10$~Gyr, which is comparable to the one determined for the MS star from a comparison with isochrones ($6.4\pm 0.7$~Gyr by \citealt{Tucci2016}; the same value but with a lower error bar of 0.3 Gyr has been obtained by \citealt{Spina2018}; the value obtained by our analysis is $7.4\pm 2.0$~Gyr). This completely resolves the age discrepancy found by \citet{Matthews2014} and \citet{bacchus2017}. We remark that the solution for the conundrum is due to the different NIR photometry we obtained from our SPHERE data, with the WD brighter at shorter wavelengths, leading to a higher temperature and a shorter cooling time; it is also due to the longer distance given by Gaia with respect to Hipparcos, which makes the MS brighter and then older.

With these values, the total mass of the HD~114174 system is of $1.73\pm 0.07$~$M_\odot$, which is much lower than what should have been the original mass (roughly 3.7-4.6~$M_\odot$, summing the mass of the progenitor of the WD and that of the MS star, or slightly less depending on the initial mass of the current MS star if this has accreted a significant amount of mass from the WD progenitor). This suggests that the system was originally closer. A detailed evolutionary model for this system would be welcomed.

The masses for the progenitors of HD~2133B and CD-56~7708B are smaller. The values of WD masses obtained by \citet{Joyce2018} correspond to progenitors with masses $<1.37$~$M_\odot$ and of $1.20\pm 0.42$~$M_\odot$, respectively, using the initial-final mass calibration by \citet{El-Badry2018}; slighty lower values were obtained using the relations by \citet{Cummings2018}. The value for CD-56~7708B obtained by our analysis is $1.8\pm 0.5$~$M_\odot$, while the mass we obtained for HD~2133B is below that expected for WD resulting from the evolution of single stars. Since these WDs are both very young, the age of the MS stars should be essentially coincident with the lifetime of the progenitors of the WDs to derive the initial mass of their progenitors. We used the STEV interface of Parsec models \citep{Bressan2012}\footnote{http://stev.oapd.inaf.it/cgi-bin/cmd} to obtain masses of $1.28^{+0.13}_{-0.08}$ and $1.65^{+1.58}_{-0.31}$~$M_\odot$ for the progenitors of HD~2133B and CD-56~7708B, respectively. Here, we neglected the WD cooling time, which is very small for these two systems. There is then substantial agreement between these different estimates: The mass of the progenitor of HD~2133B is small, likely $<1.3$~$M_\odot$ (the system should then have been in origin a nearly equal mass one), while that of the progenitor of CD-56~7708B should have been in the range of 1.4-1.8~$M_\odot$
.
\subsection{Spin-orbit alignment}

Since the rotation of the MS star is boosted by the transfer of angular momentum from the binary orbit to the star, we expect that star rotation might be aligned with the orbital one, though this may actually depend on details as to how the mass is accreted because only a very small fraction of the orbital angular momentum is actually transferred (see \citealt{Jeffries1996}). To verify this concept, we may compare the rotational periods derived from TESS with the projected rotational velocity $v~\sin{i}$ obtained from spectroscopy. If the stellar radius is known (see Table~\ref{tab:ages}), we can extract the inclination of the rotation axis of the star. 
When we apply this procedure, it is clear that HD~114174 and CD-56~7708 should be observed at a very high inclination, that is, roughly $i\sim 90$~degrees, while HD~2133 should be seen with $i\sim 30$~degrees. We found that the orbits of both HD~114174 and CD-56~7708 are indeed seen close to edge-on, which is consistent with being aligned with the orbital axis, while astrometric data are not adequate to constrain the orbit of HD~2133.

\subsection{Abundances and primary pollution}

In this subsection, we discuss the abundances of the MS stars of our systems in the context of predictions of AGB models. First, we notice that according to the model by \citet{Jeffries1996}, an effective spin-up of the MS star requires accretion of a few hundredths of solar masses of material, which is   a very small fraction of the mass lost by the AGB stars (see also \citealt{Matrozis2017}). We should compare this quantity with the mass of the outer convective envelopes of the stars under consideration that have a pristine composition that then dilutes the AGB ejecta. Using the values computed by \citet{Pinsonneault2001}, we expect that the mass of the outer convective envelope is between 0.02 and 0.03~$M_\odot$ for the program stars (with the largest value for CD-56~7708 and the smallest one for HD~2133), which is smaller or at most of the same order as the accreted mass. We then expect that dilution is not very large and that the surface abundances of the MS stars should essentially reflect that of the polluting material.

On the other hand, detailed computations are required to clarify when most of the mass transfer on the MS star occurs during the evolution of the progenitor of the WD along the AGB. In fact, \citet{Jeffries1996} showed that the accretion critically depends on the wind velocity and system separation. System separation and mass loss rates should change during the AGB evolution, so that we expect that the accretion on the MS also changes during this phase. It is then possible that the accreted material has a composition that is not identical to that at the end of the AGB phase. This complicates the comparison between nucleosynthesis models and observations.

In the case of HD~114174, the rather large mass of the WD and then for its progenitor should yield an imprint on the expected nucleosynthesis and then composition of the material transferred to the MS star. The progenitor of the WD should have had only a limited number of thermal pulses while on the AGB (see e.g., \citealt{Pastorelli2020}) because of the rather fast mass loss and evolution. A star with an initial mass of $3.2\pm 0.4$~$M_\odot$, which is as expected for the progenitor of HD~114174B, is at the top edge for becoming a C-star along the AGB (see e.g., \citealt{Abia2020, Pastorelli2020}) and for the same reason should have only a moderate production of n-capture elements through the main $s-$process, but the production critically depends on the exact mass that is assumed and on model details. We found that HD~114174A have a solar [C/Fe] ratio (=0.01$\pm$0.07 dex), and there is a modest enhancement in s-process elements Y and Ba, being [Y/Fe]=0.35$\pm$0.11 dex and [Ba/Fe]=0.20$\pm$0.10 dex, respectively (see Section~\ref{sec:primaries} for details). We exploited the FRUITY database\footnote{\url{http://fruity.oa-teramo.inaf.it}} \citep{cristallo2016} in order to compare carbon and heavy element abundances of this star to the ejecta for a 3 $M_\odot$ AGB model, with a solar metallicity and standard $^{13}$C pocket. These models produce significant amounts of C, Y, and Ba. On the other hand, the observed composition of HD~114174A nicely matches the expectations of the same models for a 4 $M_\odot$ star, but also the composition of the ejecta of a 3 $M_\odot$ AGB after only four thermal pulses. According to the model, this star is expected to have 13 thermal pulses before ending its AGB evolution.\footnote{Alternatively, one might try to derive the progenitor masses for the WDs by matching the observed composition of the MS stars with the expected composition of the ejecta of AGB stars of different masses. This approach is described in the dissertation thesis by Pacheco that can be retrieved at the URL  \url{https://teses.usp.br/teses/disponiveis/14/14131/tde-17102019-113346/pt-br.php}, where the interested reader may find more information. She found that the progenitors of the three WDs should roughly have had masses of $\sim 1.5$, $\sim 2.5$, and $\sim 2~M_\odot$ for HD~2133, HD~114174, and CD-56~7708, respectively, which is in quite good agreement with our determinations.}  

We may then consider three ways to solve this issue. First, the WD progenitor might have been more massive than we derived in our analysis; in this case, it is likely that the WD mass should also be higher. Second, it is possible that the FRUITY model overpredicts the number of thermal pulses for a 3 $M_\odot$ star: This is not excluded because the mass loss from AGB stars is not known with a high accuracy, and the upper limit for the generation of AGB stars in these models (somewhere in the range between 3 and 4 $M_\odot$) is possibly slightly larger than the value derived from observations (3 $M_\odot$: \citealt{Abia2020, Pastorelli2020}). Unfortunately, the model grid does not have intermediate masses between 3 and 4 $M_\odot$ to establish this point. Third, the accretion of mass on the MS star may have occurred during the early part of the thermal pulse phase, the binary separation (or wind speed) becoming too large for significant accretion during the very late phases of the AGB. Of course, a combination of all of these different factors may also work.

The masses of the WDs in the HD~2133 and CD-56~7708 systems are much smaller. The progenitor masses are then close to the lower edge for activation of the thermal pulses in population I stars, which is expected to be somewhere between 1.3 and 1.5~$M_\odot$ according to the FRUITY models. We then expect very low if any overabundance of C and neutron capture elements for HD~2133A, while there might be a quite substantial overabundance of these elements in CD-56~7708A if the WD, and progenitor, mass is $>1.5$~$M_\odot$, as suggested by our analysis and thoroughly consistent with the result by \citet{Joyce2018}. With respect to these predictions, we found the expected overabundance of the n-capture elements in CD-56~7708, while the [C/Fe] is subsolar (or at most solar). This difference between the behavior of C and neutron-capture elements is not predicted by models, and since it is observed in unevolved low mass stars, it cannot be attributed to the effect of the first or second dredge-up within the MS star.

Looking into the literature, we found that this different behavior is observed in other dwarf (i.e., MS) Ba-stars. For instance, \citet{Shejeelammal2020} present the case of HD~94518, which has an overabundance of $s-$process elements of 0.77~dex.\ This is very similar to what we found for CD-56~7708, but  they did not find a C-excess. They suggest that in that case, the progenitor of the WD had an initial mass of $\sim 1.5$~$M_\odot$, which is quite similar to what we found for the progenitor of CD-56~7708B.\ However, this suggestion is only based on the nucleosynthesis and not on the characteristics of the WD that was not directly observed. Cases where the mass of the WD was determined independently of nucleosynthesis are more meaningful here. Among those stars, an even more extreme case is represented by the HD~50264 system. In this case, the mass of the WD is $0.60\pm 0.05$~$M_\odot$ \citep{Escorza2019}, suggesting a progenitor with a mass of $2.0\pm 0.5$~$M_\odot$. The chemical analysis by \citet{Purandardas2019} indicates a high abundance of s-process elements of [s/Fe]$\sim 1.5$~dex, but a nearly solar C/O ratio. \citet{Allen2006a} and \citet{Allen2006b} determined the abundances of CNO and $s-$elements in 26 Ba stars, some of them being on the MS and with an independent WD mass determination. They found significantly over-solar [C/Fe] ratios in a number of them, but they were all metal-poor stars, and possibly with quite massive WD companions. HD~123585 is indeed C-rich, but it is metal poor ([Fe/H]$\sim -0.6$) and the WD mass is of $0.66\pm 0.11$~$M_\odot$ \citep{Escorza2019}, corresponding to a progenitor of $2.7\pm 0.9$~$M_\odot$. The WD companion of the more metal-rich dwarf Ba-star HD~89948 ([Fe/H]$\sim -0.4$) likely has a smaller mass of $0.54\pm 0.03$~$M_\odot$ \citep{Escorza2019}, corresponding to a progenitor of $1.4\pm 0.3$~$M_\odot$. However, in this case, the MS star has a solar [C/Fe] ratio. Furthermore, the orbital parameter distributions of dwarf Ba-stars suggests that the WD companions have a typical mass of 0.6~$M_\odot$ \citep{Escorza2019}, also indicating low-mass AGB progenitors. This might suggest that the ejecta of metal-rich and low-mass AGB stars are rich in elements produced by the $s-$process but not in C; this is not predicted by models such as FRUITY, which perhaps do not correctly reproduce dredge-up and mass loss of elements produced in different phases of the thermal pulse cycles. 

Alternatively, it is possible that the different behavior between C and $s-$process elements is due to selective accretion on the MS stars (with $s-$elements but not C accreted) or systematic differences between evolutionary scenarios (e.g., Roche-lobe overfilling versus wind). In regards to the latter, it might be interesting to note that dwarf carbon stars typically have mean periods of $<1$~year (see \citealt{Roulston2019}), while typical periods for MS Ba-stars are longer (1-100 yrs: \citealt{Escorza2019}), suggesting a different evolutionary scenario. CD-56~7708 has an even longer period. A better determination of the mass and orbit of the WDs, in particular for CD-56~7708 and HD~114174 (using e.g., further astrometry, photometry, and spectra at a shorter wavelength than possible with SPHERE IFS, e.g., using ZIMPOL at SPHERE or SHARK-VIS) as well as a systematic analysis of more Sirius-like systems similar to the one presented in this paper could help to provide more
stringent constraints on the models.

Finally, we should remind readers that there is evidence that not all of the MS stars in Sirius-like systems accreted a significant amount of mass from their pre-WD companions. This is exemplified by the case of Procyon, which has a period of about 40 yr that is shorter than those of the systems considered in this paper. The mass of Procyon~B ($0.602\pm 0.015$~$M_\odot$: \citealt{Girard2000}) implies a progenitor mass of about $2.0\pm 0.2$~$M_\odot$ using the calibration by \citet{El-Badry2018} (for comparison, \citet{Kervella2004} proposed a mass of 2.5~$M_\odot$). Such a massive AGB star should have efficiently produced C and s-process elements (see e.g., results obtained using FRUITY). However, the composition of Procyon~A is essentially indistinguishable from that of a normal MS star \citep{Kato1982, Cowley2020}, in spite of the fact that the outer convective region should be very tiny and hence dilution of accreted material should be very small. After noticing this fact, \citet{North2020} recently proposed that the lack of overabundances of C and $s-$elements can be related to anticorrelations between their enhancement with period and metallicity in Ba-stars. However, CD-56~7708 is much more extreme than Procyon in both of these respects, but it still has a significant excess of $s-$elements. Further work is clearly required to understand the details of the Ba-stars phenomenon. 

\begin{acknowledgements}
We thank Franca D'Antona and Paolo Ventura for useful comments. TAP and JM thank Diego Lorenzo-Oliveira for help with processing of the HARPS/FEROS spectra.
SPHERE is an instrument designed and built by a consortium consisting of IPAG (Grenoble, France), MPIA (Heidelberg, Germany), LAM (Marseille, France), LESIA (Paris, France), Laboratoire Lagrange (Nice, France), INAF - Osservatorio di Padova (Italy), Observatoire de Gen\`eve (Switzerland), ETH Z\"urich (Switzerland), NOVA (Netherlands), ONERA (France) and ASTRON (Netherlands) in collaboration with ESO. SPHERE was funded by ESO, with additional contributions from CNRS (France), MPIA (Germany), INAF (Italy), FINES (Switzerland) and NOVA (Netherlands). SPHERE also received funding from the European Commission Sixth and Seventh Framework Programmes as part of the Optical Infrared Coordination Network for Astronomy (OPTICON) under grant number RII3-Ct-2004-001566 for FP6 (2004--2008), grant number 226604 for FP7 (2009--2012) and grant number 312430 for FP7 (2013--2016).
This work has been supported by the project PRIN INAF 2016 The Craddle of Life - GENESIS-SKA (General Conditions in Early Planetary Systems for the rise of life with SKA) and by the "Progetti Premiali" funding scheme of the Italian Ministry of Education, University, and Research.  A.Z. acknowledges support from the FONDECYT Iniciaci\'on en investigaci\'on project number 11190837. JM thanks FAPESP (2018/04055-8). This study financed in part by the Coordena\c{c}\~ao de Aperfei\c{c}oamento de Pessoal de Nível Superior - Brasil (CAPES) - Finance Code 001. C. P. acknowledge financial support from Fondecyt (grant 3190691) and financial support from the ICM (Iniciativa Cient\'ifica Milenio) via the N\'ucleo Milenio  de  Formaci\'on Planetaria grant, from the Universidad de Valpara\'iso.
We have made use of the GPI/Gemini observation under proposal ID GS-2018A-FT-103 (PI Pacheco).
This research has made use of the SIMBAD database and Vizier services, operated at CDS, Strasbourg, France and of the Washington Double Star Catalog maintained at the U.S. Naval Observatory. 
This work has made use of data from the European Space Agency (ESA) mission {\it Gaia} (\url{https://www.cosmos.esa.int/Gaia}), processed by the {\it Gaia} Data Processing and Analysis Consortium (DPAC, \url{https://www.cosmos.esa.int/web/Gaia/dpac/consortium}). Funding for the DPAC has been provided by national institutions, in particular the institutions participating in the {\it Gaia} Multilateral Agreement.
This paper includes data collected with the TESS mission, obtained from the MAST data archive at the Space Telescope Science Institute (STScI). Funding for the TESS mission is provided by the NASA Explorer Program. STScI is operated by the Association of Universities for Research in Astronomy, Inc., under NASA contract NAS 5–26555.
This paper has made use of data products available in ESO archive. Program ID: 188.C-0265 (PI: Mel\'endez),  089.D-0097 (PI: Helminiak), 087.D-0012, (PI: Helminiak)

\end{acknowledgements}

\bibliographystyle{aa} 
\bibliography{main}

\section{Appendix: Radial velocities from HARPS data for HD114174}

This appendix reports on the radial velocities for HD~114174 obtained from the HARPS data (see Table~\ref{tab:my_label}). The values we considered were obtained using the HARPS pipeline, using a G2 mask, as available on the reduced data archive\footnote{ 
\url{http://archive.eso.org/wdb/wdb/adp/phase3_main/form}}. In the analysis, we corrected the radial velocities for the offsets with respect to the Keck radial velocities \citep{crepp2013, Butler2017} using the overlapping epochs.

\begin{table}[htb]
    \caption{Radial velocities from HARPS data.}
    \centering
    \begin{tabular}{lcc}
\hline
\hline
BJD       &  V$_r$ (km/s) &  Err (m/s)\\
\hline
2455983.809     &       24.91643        &       0.39    \\
2455983.813     &       24.91563        &       0.37    \\
2455984.794     &       24.91685        &       0.46    \\
2455984.798     &       24.91688        &       0.43    \\
2455985.784     &       24.91449        &       0.38    \\
2455985.789     &       24.91446        &       0.38    \\
2455986.797     &       24.91795        &       0.45    \\
2455986.802     &       24.91540        &       0.42    \\
2456047.656     &       24.93232        &       0.53    \\
2456047.662     &       24.93264        &       0.50    \\
2456048.648     &       24.93120        &       0.48    \\
2456048.653     &       24.93207        &       0.51    \\
2456298.886     &       24.96255        &       0.43    \\
2456298.891     &       24.98727        &       0.43    \\
2456300.842     &       24.95480        &       0.50    \\
2456300.847     &       24.95483        &       0.44    \\
2456301.821     &       24.95537        &       0.56    \\
2456301.826     &       24.95748        &       0.51    \\
2456375.745     &       24.96535        &       0.41    \\
2456375.750     &       24.96817        &       0.42    \\
2456376.734     &       24.97047        &       0.46    \\
2456376.739     &       24.97039        &       0.46    \\
2456376.745     &       24.97050        &       0.47    \\
2456376.750     &       24.97092        &       0.47    \\
2456376.756     &       24.96910        &       0.45    \\
2456376.761     &       24.97002        &       0.44    \\
2456377.713     &       24.97116        &       0.40    \\
2456377.718     &       24.96900        &       0.38    \\
2456378.701     &       24.97094        &       0.40    \\
2456378.706     &       24.96883        &       0.40    \\
2456379.725     &       24.96903        &       0.41    \\
2456379.730     &       24.96929        &       0.39    \\
2456380.730     &       24.97143        &       0.36    \\
2456380.735     &       24.97146        &       0.36    \\
2456381.728     &       24.97021        &       0.37    \\
2456381.733     &       24.97166        &       0.35    \\
2456484.508     &       24.98923        &       0.42    \\
2456484.513     &       24.98742        &       0.47    \\
2456485.511     &       24.98729        &       0.41    \\
2456485.516     &       24.98803        &       0.47    \\
2456486.497     &       24.99218        &       0.56    \\
2456486.503     &       24.98650        &       0.34    \\
2456487.497     &       24.98909        &       0.35    \\
2456487.502     &       24.98847        &       0.36    \\
2456490.492     &       24.99658        &       0.42    \\
2456490.498     &       24.99768        &       0.44    \\
2456708.767     &       25.02976        &       0.46    \\
2456708.772     &       25.02940        &       0.42    \\
2456709.788     &       25.02923        &       0.48    \\
2456709.793     &       25.02795        &       0.48    \\
2456710.786     &       25.02827        &       0.52    \\
2456710.791     &       25.02734        &       0.54    \\
2456711.759     &       25.02848        &       0.46    \\
2456711.764     &       25.02963        &       0.45    \\
2456851.519     &       25.05240        &       0.43    \\
2456851.524     &       25.05155        &       0.52    \\
2456852.482     &       25.05299        &       0.51    \\
2456852.488     &       25.05233        &       0.55    \\
2456854.487     &       25.05340        &       0.60    \\
2456854.492     &       25.05279        &       0.57    \\
\hline
\end{tabular}
    \label{tab:my_label}
\addtocounter{table}{-1}
\end{table}

\begin{table}[htb]
\caption{Cont...}
    \centering
    \begin{tabular}{lcc}
\hline
\hline
JD       &  V$_r$ (km/s) &  Err (km/s)\\
\hline
2456855.475     &       25.05160        &       0.37    \\
2456855.480     &       25.04988        &       0.38    \\
2456856.479     &       25.05329        &       0.47    \\
2456855.475     &       25.05160        &       0.37    \\
2456855.480     &       25.04988        &       0.38    \\
2456856.484     &       25.05405        &       0.44    \\
2457025.847     &       25.07921        &       0.38    \\
2457025.852     &       25.07853        &       0.37    \\
2457230.500     &       25.12578        &       0.39    \\
2457230.505     &       25.12560        &       0.39    \\
2457505.666     &       24.90305        &       0.24    \\
2457505.671     &       24.90649        &       0.23    \\
2457764.872     &       25.20305        &       0.30    \\
2457764.877     &       25.20412        &       0.30    \\
2457807.828     &       25.21461        &       0.39    \\
2457807.833     &       25.21429        &       0.88    \\
2457808.796     &       25.21086        &       0.43    \\
2457808.801     &       25.21080        &       0.41    \\
\hline
\hline    
    \end{tabular}
\end{table}

\end{document}